\documentclass[manuscript,screen]{acmart}
\pdfoutput=1

\setcopyright{none}

\usepackage[utf8]{inputenc}
\usepackage{array}
\usepackage{algorithm}
\usepackage{algpseudocode}
\usepackage{amsmath}
\usepackage{amssymb}
\usepackage[english]{babel}
\usepackage{enumitem}
\usepackage{hyperref}
\usepackage{cleveref}
\usepackage[caption=false]{subfig}
\usepackage{tikz}
\usepackage{pgfplots}
\usepackage{verbatim}

\acmJournal{TOMS}
\title[Exploiting structure: optimal code generation for high-order
FEM]{Exposing and exploiting structure: optimal code generation for
  high-order finite element methods}

\author{Miklós Homolya}
\affiliation{%
  \institution{Imperial College London}
  \department{The Grantham Institute}
  \department{Department of Computing}
  \streetaddress{South Kensington Campus}
  \city{London}
  \postcode{SW7 2AZ}
  \country{UK}}
\email{m.homolya14@imperial.ac.uk}

\author{Robert C. Kirby}
\affiliation{%
  \institution{Baylor University}
  \department{Department of Mathematics}
  \streetaddress{One Bear Place}
  \city{Waco}
  \state{TX}
  \country{USA}}
\email{robert_kirby@baylor.edu}

\author{David A. Ham}
\orcid{0000-0001-9545-9110}
\affiliation{%
  \institution{Imperial College London}
  \department{Department of Mathematics}
  \streetaddress{South Kensington Campus}
  \city{London}
  \postcode{SW7 2AZ}
  \country{UK}}
\email{david.ham@imperial.ac.uk}

\numberwithin{equation}{section}

\DeclareMathOperator{\tr}{tr}

\crefname{algorithm}{Algorithm}{Algorithms}
\crefname{figure}{Fig.}{Figs.}
\crefname{table}{Table}{Tables}

\newcommand{\dx}{\, \mathrm{d}x}
\newcommand{\dX}{\, \mathrm{d}\hat{x}}

\newcommand{\Hdiv}{$ H( \mathrm{div} ) $}
\newcommand{\Hcurl}{$ H( \mathrm{curl} ) $}

\begin{abstract}
  Code generation based software platforms, such as Firedrake, have
  become popular tools for developing complicated finite element
  discretisations of partial differential equations.  We extended the
  code generation infrastructure in Firedrake with optimisations that
  can exploit the structure inherent to some finite elements.  This
  includes sum factorisation on cuboid cells for continuous,
  discontinuous, \Hdiv{} and \Hcurl{} conforming elements. Our
  experiments confirm optimal algorithmic complexity for high-order
  finite element assembly.  This is achieved through several novel
  contributions: the introduction of a more powerful interface between
  the form compiler and the library providing the finite elements; a
  more abstract, smarter library of finite elements called FInAT that
  explicitly communicates the structure of elements; and form compiler
  algorithms to automatically exploit this exposed structure.
\end{abstract}



\thanks{This work was supported by \grantsponsor{grantham}{The
    Grantham Institute}{http://www.imperial.ac.uk/grantham/}; the
  \grantsponsor{nsf}{National Science
    Foundation}{https://www.nsf.gov/} [grant number
  \grantnum{nfs}{1525697}]; and the \grantsponsor{nerc}{Natural
    Environment Research Council}{http://www.nerc.ac.uk/} [grant
  number \grantnum{nerc}{NE/K008951/1}].}

\begin{document}

\maketitle

\section{Introduction}

Code generation based software tools have become an increasingly popular
mechanism for the implementation of complicated numerical solvers for partial
differential equations using the finite element method.  Examples of
such software platforms include FreeFem++~\cite{Hecht2012},
FEniCS \cite{logg2012automated,Alnaes2015}, and
Firedrake~\cite{Rathgeber2016}.  These software packages enable
high-productivity development of reasonably efficient numerical codes.
They are most commonly used for low-order finite element
discretisations of complicated PDEs, while code  for the assembly of high-order finite element
discretisations has been predominantly by manually written.

High-order finite element methods, such as the spectral element
method~\cite{Patera1984}, combine the convergence properties of
spectral methods with the geometric flexibility of traditional linear
FEM.  Because they provide a larger amount of work per
degree-of-freedom, they are also better suited to make use of modern
hardware architectures, which exhibit a growing gap between processor and memory
speeds.  These techniques have been successfully employed in a number
of research packages: NEK5000~\cite{nek5000},
Nektar++~\cite{Cantwell2015} and PyFR~\cite{Witherden2014} for
computational fluid dynamics, HERMES~\cite{Vejchodsky2007} for
Maxwell's equations, EXA-DUNE~\cite{Bastian2014} for porous media
flows, pTatin3D~\cite{May2014} for lithospheric dynamics.
\citet{Hientzsch2001} implemented an efficient discretisation of
Maxwell's equations using \Hcurl{} conforming spectral elements.
However, according to \citet{Cantwell2015}, the implementational
complexity of these techniques has limited their uptake in many
application domains, especially outside academia.

FEniCS and Firedrake depend on the \emph{FInite element Automatic Tabulator}
(FIAT) \cite{Kirby2004}, a library of finite elements. Since FIAT generally
supports finite elements with arbitrary order, FEniCS and Firedrake do in
principle support high-order finite element discretisations.  However, their
implementations have hitherto lacked the optimisations that are crucial at
high order, resulting in much slower performance than hand-written spectral
element codes. The reason that this deficiency has not been corrected in any
of the previous work on improved code generation in FEniCS and Firedrake 
\cite{Kirby2007,Oelgaard2010,Luporini2015,Luporini2016,Homolya2017} is that
the interface which FIAT presents to form compilers prevents the
exploitation of the required optimisations.

To implement the necessary optimising transformations, one needs to
rearrange finite element assembly loops in ways that take into account
the structure inherent to some finite elements.  FIAT, however, does
not and cannot express such structure, because its interface provides
no way to do so.  We therefore hereby present FInAT, a smarter
library of finite elements that can express structure within elements
by introducing a novel interface between the form compiler and the
finite element library.

The rest of this paper is arranged as follows.  In the remainder of
this section we review the relevant steps of finite element assembly,
describe the limitations of FIAT in more details, and finally list the
novel ideas incorporated in this work.  \Cref{sec:finat} introduces
FInAT and its structure-preserving element implementations.  \Cref{sec:tsfc}
describes the form compiler algorithms implemented in the
\emph{Two-Stage Form Compiler} (TSFC) \cite{Homolya2017} that exploit
the structure of elements to optimise finite element kernels.  We
verify the results through experimental evaluation in
\cref{sec:evaluation}, and \cref{sec:conclusion} concludes the paper.

\subsection{Background}
\label{sec:background}

We start with a brief recap of finite element assembly to show how
code generation based solutions (e.g., FEniCS and Firedrake) make use
of a finite element library such as FIAT.  Consider the stationary
heat equation $ -\nabla \cdot (\kappa \nabla u) = f $ with $ \kappa $
thermal conductivity and $ f $ heat source on a domain $ \Omega $ with
$ u = 0 $ on the boundary.  Its standard weak form with left-hand side
$ a(u, v) $ and right-hand side $ L(v) $ is given by
\begin{align}
  \label{eq:weighted_laplace}
  a(u,v) &= \int_\Omega \kappa \nabla u \cdot \nabla v \dx \quad \text{and} \\
  L(v) &= \int_\Omega f v \dx.
\end{align}
Such weak forms are defined in code using the \emph{Unified Form
  Language} (UFL) \cite{Alnes2014}.  The prescribed spatial functions
are often represented as functions from finite element function
spaces.  In UFL terminology, $ v $ and $ u $ are called
\emph{arguments}, and $ \kappa $, and $ f $ are called
\emph{coefficients} of the multilinear forms $ a $ and $ L $.

For a simpler discussion, we take $ \kappa \equiv 1 $, so the
left-hand side reduces to the well-known Laplace operator:
\begin{equation}
  \label{eq:laplace_form}
  a(u, v) = \int_\Omega \nabla u \cdot \nabla v \dx .
\end{equation}
Following the notation in \citet{Kirby2014}, let $ \mathcal{T} $
be a tessellation of the domain $ \Omega \subset \mathbb{R}^d $, so
the integral can be evaluated cellwise:
\begin{equation}
  \int_\Omega \nabla u \cdot \nabla v \dx =
  \sum_{K \in \mathcal{T}} \int_{K} \nabla u \cdot \nabla v \dx .
\end{equation}

Suppose we have a reference cell $ \hat{K} $ such that each
$ K \in \mathcal{T} $ is diffeomorphic to $ \hat{K} $ via a mapping
$ F_K : \hat{K} \rightarrow K $.  Let
$ {\left. u \right\vert}_K : K \rightarrow \mathbb{R} $ be the
restriction of $ u $ to cell $ K $, and
$ \hat{u}_K = {\left. u \right\vert}_K \circ F_K^{-1} $ its pullback to the
reference cell $ \hat{K} $.  This implies
\begin{align}
  {\left. u \right\vert}_K &= \phantom{J_K^{-T} \hat{\nabla}}
                             \hat{u}_K \circ F_K^{-1} \\
  {\left. \nabla u \right\vert}_K &= J_K^{-T} \hat{\nabla} \hat{u}_K
                                    \circ F_K^{-1}
\end{align}
where $ \hat{\nabla} $ indicates differentiation in reference
coordinates, and $ J_K = \hat{\nabla} F_K $ is the Jacobian matrix.
Transforming the integral from physical to reference space, we have
\begin{equation}
  \int_K \nabla u \cdot \nabla v \dx
  = \int_{\hat{K}} J_K^{-T} \hat{\nabla} \hat{u}_K \cdot J_K^{-T}
    \hat{\nabla} \hat{v}_K \left| J_K \right| \dX .
\end{equation}

Let us consider the evaluation of these integrals via numerical
quadrature.  Let $ \{ \xi_q \}_{q=1}^{N_q} $ be a set of quadrature
points on $ \hat{K} $ with corresponding quadrature weights
$ \{ w_q \}_{q=1}^{N_q} $, so the integral is approximated by
\begin{equation}
  \label{eq:laplace_quadrature}
  \sum_{q=1}^{N_q} w_q ( J_K^{-T}(\xi_q) \hat{\nabla}
  \hat{u}_K(\xi_q) ) \cdot ( J_K^{-T}(\xi_q) \hat{\nabla}
  \hat{v}_K(\xi_q) ) \left| J_K(\xi_q) \right| .
\end{equation}

Now let $ \{ \Psi_i \}_{i=1}^{N_f} $ be a reference basis, and let
$ \hat{u}_K $ and $ \hat{v}_K $ be expressed in this basis.  Using the
abbreviations $ G_q := {\left[ J_K(\xi_q) \right]}^{-1} $ and
$ S_q := \left| J_K(\xi_q) \right| $, the element stiffness matrix is
then evaluated as
\begin{equation}
  \label{eq:laplace_prefiat}
  A^K_{ij} =
  \sum_{q=1}^{N_q} w_q S_q ( G_q^T \hat{\nabla} \Psi_i(\xi_q) ) \cdot
  ( G_q^T \hat{\nabla} \Psi_j(\xi_q) ) .
\end{equation}

Given the weak form as in \cref{eq:laplace_form}, a form compiler such
as TSFC carries out the above steps automatically.
FIAT provides tables of basis functions and their derivatives at quadrature
points, such as the numerical tensor $ D\mathbf{\Psi} $,
\begin{equation}
  \label{eq:fiat_table}
  D\mathbf{\Psi}_{iqk} := \frac{\partial \Psi_i}{\partial \hat{x}_k}(\xi_q) ,
\end{equation}
so that one could substitute
$ \hat{\nabla} \Psi_i(\xi_q) \mapsto D\mathbf{\Psi}_{iq} $ and
$ \hat{\nabla} \Psi_j(\xi_q) \mapsto D\mathbf{\Psi}_{jq} $.  With these
substitutions, and assuming that we know how to evaluate $ J_K(\xi_q) $,
\cref{eq:laplace_prefiat} becomes a straightforward tensor algebra
expression.  For further steps towards generating C code, we refer the
reader to \citet[\S 4]{Homolya2017}.
Finally, the assembled local matrix $ A^K $ is added to the global
sparse matrix $ A $; this step is known as \emph{global assembly}.

The assembled global matrix can be defined as
\begin{equation}
  A_{ij} = a(\psi_i, \psi_j)
\end{equation}
where $ \{ \psi_i \}_{i=1}^{N_g} $ is the global basis.  Large systems
of linear equations, such as those resulting from finite element
problems, are generally solved using various \emph{Krylov subspace
  methods}.  These methods do not strictly require \emph{matrix
  assembly}, as they only directly use the action of the linear
operator.  (Many preconditioners and direct solvers require matrix entries of
the assembled operator, however.)  Instead of assembling the operator as a
sparse matrix, and then applying matrix-vector multiplications, one
can also assemble the operator action on vector $ U $ directly as a
parametrised linear form:
\begin{align}
  (A U)_i &= \sum_{j=1}^{N_g} A_{ij} U_j = \sum_{j=1}^{N_g} a(\psi_i, \psi_j) U_j \\
          &= a(\psi_i, \sum_{j=1}^{N_g} U_j \psi_j) = a(\psi_i, u)
\end{align}
where $ u(x) = \sum_{j=1}^{N_g} U_j \psi_j(x) $ is a function
isomorphic to the vector $ U $.  This approach is commonly known as a
\emph{matrix-free method}.

\subsection{Limitations}
\label{sec:limits}

The approach of using FIAT as in \cref{eq:fiat_table} enables us to
generate code for matrix assembly,
however, it also renders certain optimisations
infeasible which rely on the structure inherent to some finite
elements.  This is because FIAT does not, and cannot express such
structure, since the tabulations it provides are just numerical
arrays, henceforth called \emph{tabulation matrices}.

For example, \emph{sum factorisation} is a well-known technique that
drastically improves the assembly performance of high-order
discretisations.  It relies on being able to write tabulation matrices
as a tensor product of smaller matrices.  Suppose the reference cell
is a square.  We take one-dimensional quadrature rules
$ \{ (\xi^{(1)}_{q_1}, w^{(1)}_{q_1}) \}_{q_1=1}^{N_{q_1}} $ and
$ \{ (\xi^{(2)}_{q_2}, w^{(2)}_{q_2}) \}_{q_2=1}^{N_{q_2}} $ to define
their tensor product with
\begin{align}
  \label{eq:tpps}
  \xi_q &= \begin{bmatrix}
    \xi^{(1)}_{q_1} \\ \xi^{(2)}_{q_2}
  \end{bmatrix} \qquad \text{and}
  \\
  w_q &= w^{(1)}_{q_1} w^{(2)}_{q_2}
\end{align}
where $ q = (q_1, q_2) $.  Similarly, we take one-dimensional finite
element bases $ \{ \Psi^{(1)}_{i_1} \}_{i_1=1}^{N_{f_1}} $ and
$ \{ \Psi^{(2)}_{i_2} \}_{i_2=1}^{N_{f_2}} $ to define the tensor
product element $ \{ \Psi_i \}_i $ such that
\begin{equation}
  \label{eq:limits_tpe}
  \Psi_i(\hat{x}) = \Psi^{(1)}_{i_1}(\hat{x}_1) \Psi^{(2)}_{i_2}(\hat{x}_2)
\end{equation}
where $ i = (i_1, i_2) $, and $ \hat{x} = (\hat{x}_1, \hat{x}_2) $.
Consequently,
$ \Psi_i(\xi_q) = \Psi^{(1)}_{i_1}( \xi^{(1)}_{q_1})
\Psi^{(2)}_{i_2}(\xi^{(2)}_{q_2}) $, so the relationship between
tabulation matrices is
\begin{equation}
  \mathbf{\Psi}_{i q} = \mathbf{\Psi}^{(1)}_{i_1 q_1} \mathbf{\Psi}^{(2)}_{i_2 q_2} .
\end{equation}
In other words, the tabulation matrix of the tensor product element is
the tensor -- or, more specifically, Kronecker -- product of the
tabulation matrices of the one-dimensional
elements.  We will see later how this structure can be exploited for
performance gain, however, it shall be clear that FIAT can only
provide $ \Psi_{i q} $ as a numerical matrix, and cannot express this
product structure.

Vector elements, such as used when each component of a fluid velocity
or elastic displacement is expressed in the same scalar space, are
another example.  Suppose 
$ \{ \Psi^*_i \}_{i=1}^{N_{f^*}} $ is a scalar-valued finite element
basis on $ \hat{K} $.  We can define the vector-valued version of this
element with basis functions
\begin{equation}
  \Psi_i = \Psi^*_{i_1} \mathbf{\hat{e}}_{i_2}
\end{equation}
where $ i = (i_1, i_2) $, and $ \mathbf{\hat{e}}_{i_2} $ is the unit
vector in the $ i_2 $-th direction.  The tabulation matrices of the
scalar and the vector element are related: we get the $ k $-th
component of the $ i $-th basis function of the vector element at the
$ q $-th quadrature point as
\begin{equation}
  \mathbf{\Psi}_{i q k} = \mathbf{\Psi}^*_{i_1 q} \delta_{i_2 k}
\end{equation}
where $ \delta_{i_2 k} $ denotes the \emph{Kronecker delta} (which is
one for $ i_2 = k $ and zero otherwise).  FIAT cannot express
this structure.  Instead, for each component $ k $, FIAT provides a
tabulation matrix with zero blocks for those basis functions which do
not contribute to the $ k $-th component.  As a result, tabulation
matrices are $ d $-fold larger than they need to be, wasting both
static storage and run-time operations.
The latter can even increase $ d^2 $-fold for matrix assembly.

To conclude this section, the calculations of bilinear forms such as
\cref{eq:laplace_prefiat} hitherto incorporated certain tables of data from the
finite element library such as $ D\Psi_{iq} $ in (\ref{eq:fiat_table}).
However, any expression of $ i $ and $ q $ that evaluates to the
proper value is also suitable, and given proper structure and compiler
transformations, can be very advantageous.  This paper contributes the
following novel ideas:
\begin{enumerate}[label={\arabic*.}]
\item Providing such evaluations as expressions (including simple
  table look-up) in a language that the form compiler understands.
\item Utilising this language to express additional structure such as
  factored basis functions.
\item Form compiler algorithms for optimising common patterns
  appearing.
\end{enumerate}
Note that 1 and 2 are proper to FInAT, while 3 is an application of
those ideas that we implemented in TSFC.

\section{\NoCaseChange{FInAT}}
\label{sec:finat}

Unlike FIAT, ``FInAT Is not A Tabulator.''  Instead, it provides
symbolic expressions for the evaluation of finite element basis
functions at a set of points, henceforth called \emph{tabulation
  expressions}.  Thus FInAT is able to express the structure that is
intrinsic to some finite elements.  This, of course, requires the
definition of an expression language for these tabulation expressions.

To facilitate integration with TSFC, tabulations are provided in the
tensor algebra language GEM \cite{Homolya2017}.  GEM originates as the
intermediate representation of TSFC, so TSFC can directly insert
tabulation expressions into partially compiled form expressions.

\subsection{Overview of GEM}
\label{sec:gem_overview}

This summary of GEM is based on \citet[\S 3.1 and \S
4.1]{Homolya2017}.  We begin with a concise listing of all node types:
\begin{itemize}
\item \textbf{Terminals:}
  \begin{itemize}
  \item \texttt{Literal} (tensor literal)
  \item \texttt{Zero} (all-zero tensor)
  \item \texttt{Identity} (identity matrix)
  \item \texttt{Variable} (run-time value, for kernel arguments)
  \end{itemize}
\item \textbf{Scalar operations:}
  \begin{itemize}
  \item \textit{Binary operators:} \texttt{Sum}, \texttt{Product},
    \texttt{Division}, \texttt{Power}, \texttt{MinValue}, \texttt{MaxValue}
  \item \textit{Unary operators:} \texttt{MathFunction} (e.g.\@ $ \sin $, $ \cos $)
  \item \texttt{Comparison} ($>$, $\ge$, $=$, $\ne$, $<$, $\le$):
    compares numbers, returns Boolean
  \item \textit{Logical operators:} \texttt{LogicalAnd},
    \texttt{LogicalOr}, \texttt{LogicalNot}
  \item \texttt{Conditional}: selects between a ``true'' and a
    ``false'' expression based on a Boolean valued \emph{condition}
   \end{itemize}
\item \textbf{Index types:}  
  \begin{itemize}
  \item \texttt{int} (fixed index)
  \item \texttt{Index} (free index): creates a loop at code generation
  \item \texttt{VariableIndex} (unknown fixed index): index value only
    known at run-time, e.g.\@ facet number
 \end{itemize}
\item \textbf{Tensor nodes:} \texttt{Indexed},
  \texttt{FlexiblyIndexed}, \texttt{ComponentTensor},
  \texttt{IndexSum}, \texttt{ListTensor} \\
  See notes below, and refer to \citet[\S 3.1]{Homolya2017} for further
  discussion.
\item \textbf{Special nodes:}
  \begin{itemize}
  \item \texttt{Delta} (Kronecker delta)
  \item \texttt{Concatenate}: vectorises each operand and concatenates
    them.
  \end{itemize}
\end{itemize}

It is important to understand that the tensor nature of GEM
expressions is represented as \emph{shape} and \emph{free indices}:
\begin{description}
\item[shape] An ordered list of dimensions and their respective
  extent, e.g.\@ (2, 2). A dimension is only identified by its
  position in the shape.
\item[free indices] An unordered set of dimensions where each
  dimension is identified by a symbolic index object.  One might think
  of free indices as an ``unrolled shape''.
\end{description}
These traits are an integral part of any GEM expression.  For example,
let $ A $ be a $ 2 \times 2 $ matrix, then $ A $ has shape (2, 2) and
no free indices. $ A_{1,1} $, written as \verb|Indexed(A, (1, 1))|,
has scalar shape and no free indices; $ A_{i,j} $, written as
\verb|Indexed(A, (i, j))|, has scalar shape and free indices $ i $ and
$ j $; and $ A_{i,1} $, written as \verb|Indexed(A, (i, 1))|, has
scalar shape and free index $ i $.

\texttt{ComponentTensor}, in some sense, is the inverse operation of
\texttt{Indexed}.  That is, if
\begin{equation}
  A = \mathtt{ComponentTensor}(e, \alpha) ,
\end{equation}
then $ A_\alpha = e $, where
$ \alpha := (\alpha_1, \alpha_2, \ldots, \alpha_k) $ is called a
multi-index.  Later in this paper, we typically use the concise
notation
\begin{equation}
  \label{eq:component_tensor}
  ]e[_\alpha \equiv \mathtt{ComponentTensor}(e, \alpha) .
\end{equation}
In order to be a well-formed expression, $ e $ must be an expression
with scalar shape and free indices
$ \alpha_1, \alpha_2, \ldots, \alpha_k $ (at least).  Then
$ ]e[_\alpha $ is a tensor with shape
$ (\alpha_1\mathtt{.extent}, \alpha_2\mathtt{.extent}, \ldots,
\alpha_k\mathtt{.extent}) $.  That is, the free indices in $ \alpha $
are made into shape.

Most supported operations, such as addition and multiplication,
naturally require scalar shape.  So operands with non-scalar shape
need indexing before most operations, but the shape can be restored by
wrapping the result in a \texttt{ComponentTensor}.

\subsection{Generic FIAT element wrapper}
\label{sec:fiat_wrapper}

Let $ \{ \xi_q \}_{q=1}^{N_q} $ be a set of (quadrature) points on the
reference cell, and $ \{ \Psi_i \}_{i=1}^{N_f} $ be the basis
functions of a reference finite element.
If FIAT implements this element, then it can produce a tabulation
matrix $ \mathbf{\Psi} $ such that
$ \mathbf{\Psi}_{i q} := \Psi_i (\xi_q) $.  Since GEM can represent
literal matrices and indexing, FInAT can invoke FIAT and construct a
trivial tabulation expression, providing the substitution:
\begin{equation}
  \label{eq:fiat_compat}
  \Psi_i (\xi_q) \mapsto \mathbf{\Psi}_{i q}
\end{equation}
and similarly for derivatives of basis functions.
This generic wrapper removes the need for the form compiler to
interface both FIAT and FInAT at the same time, while all FIAT
elements continue to be available with no regression.

\subsection{Vector and tensor elements}
\label{sec:tensor_element}

A \emph{vector element} constructs a vector-valued element by
duplicating a scalar-valued element for each component.  Fluid
velocities, e.g., are often represented using vector elements of
Lagrange elements.  Let $ \{ \Psi^*_\alpha \}_{\alpha \in A} $ be the
basis of a scalar-valued finite element.  When the basis has a
tensor-product structure,
$ \alpha = (\alpha_1, \alpha_2, \ldots, \alpha_s) $ is a multi-index.
The corresponding $ d $-dimensional vector element has basis
\begin{equation}
  \label{eq:vector_element_def}
  \Psi_{(\alpha,j)} = \Psi^*_{\alpha} \mathbf{\hat{e}}_j
\end{equation}
where $ \mathbf{\hat{e}}_j $ is the $ d $-dimensional unit vector
whose $ j $-th coordinate is $ 1 $.

The scalar element may have its own structure, so let
$ \mathcal{E}\{ \Psi^*_\alpha (\xi_q) \} $ denote its
tabulation expression at points $ \{ \xi_q \}_{q=1}^Q $.  Then FInAT
constructs the following symbolic expression for the evaluation of the
vector element:
\begin{equation}
  \label{eq:vector_element_subst}
  \Psi_{(\alpha,j) k}(\xi_q) \mapsto \mathcal{E}\{ \Psi^*_\alpha
  (\xi_q) \} \delta_{jk} .
\end{equation}

A vector element is a rank-$ 1 $ \emph{tensor element}.%
\footnote{Note the distinction between ``tensor element'' and ``tensor
  product'' element.  By \emph{tensor element} we mean a finite
  element whose function space members take on tensorial values.  By
  \emph{tensor product element} we mean that the basis functions can
  be written as a product of lower-dimensional basis functions.}%
\addtocounter{footnote}{-1}\addtocounter{Hfootnote}{-1}
FInAT can also evaluate rank-$ n $ tensor elements as
\begin{equation}
  \label{eq:tensor_element_subst}
  \Psi_{(\alpha,\nu) \kappa}(\xi_q) \mapsto \mathcal{E}\{
  \Psi^*_\alpha (\xi_q) \} \prod_{i=1}^n \delta_{\nu_i \kappa_i} .
\end{equation}

\subsection{Tensor product element}
\label{sec:product_element}

Here we follow \citet{McRae2014} for a definition of the \emph{tensor
  product} element\footnotemark.  Let $ K_1 \subset \mathbb{R}^{d_1} $
and $ K_2 \subset \mathbb{R}^{d_2} $ be reference cells, the reference
tensor product cell $ K_1 \times K_2 $ is defined as
\begin{equation}
  \label{eq:product_cell}
  K_1 \times K_2 = \left\{ (\hat{x}_1, \ldots, \hat{x}_{d_1+d_2}) \in
    \mathbb{R}^{d_1+d_2} \mid (\hat{x}_1, \ldots, \hat{x}_{d_1}) \in K_1 , (\hat{x}_{d_1
      + 1}, \ldots, \hat{x}_{d_1 + d_2}) \in K_2 \right\} .
\end{equation}
Similarly, let $ \{ \xi^{(1)}_{q_1} \}_{q_1=1}^{Q_1} $ and
$ \{ \xi^{(2)}_{q_2} \}_{q_2=1}^{Q_2} $ be quadrature points on
$ K_1 $ and $ K_2 $ respectively.  The tensor product point set
$ \{ \xi_{(q_1,q_2)} \}_{(q_1,q_2)} $ is defined as
\begin{align}
  \label{eq:product_point_set}
  & \xi_{(q_1,q_2)} = (\hat{x}_1, \ldots, \hat{x}_{d_1}, \hat{x}_{d_1
    + 1}, \ldots, \hat{x}_{d_1+d_2}) \\
  &
  \begin{aligned}
    \mbox{where }
    & (\hat{x}_1, \ldots, \hat{x}_{d_1}) = \xi^{(1)}_{q_1}, \mbox{ and} \\
    & (\hat{x}_{d_1+1}, \ldots, \hat{x}_{d_1+d_2}) = \xi^{(2)}_{q_2} .
  \end{aligned}
\end{align}
Let $ \{ \Psi^{(1)}_{i_1} \}_{i_1=1}^{N_{f,1}} $ and
$ \{ \Psi^{(2)}_{i_2} \}_{i_2=1}^{N_{f,2}} $ be finite element bases
on the reference cells $ K_1 $ and $ K_2 $ respectively.  The tensor
product element on reference cell $ K := K_1 \times K_2 $ has basis
functions
\begin{equation}
  \label{eq:product_basis}
  \Psi_{(i_1,i_2)}(\hat{x}_1, \ldots, \hat{x}_{d_1+d_2}) =
  \Psi^{(1)}_{i_1}(\hat{x}_1, \ldots, \hat{x}_{d_1}) \,
  \Psi^{(2)}_{i_2}(\hat{x}_{d_1+1}, \ldots, \hat{x}_{d_1+d_2}) .
\end{equation}

\begin{figure}
  \centering
  \begin{tabular}{m{3cm} m{1em} m{1em} m{1em} m{3cm}}
    \includegraphics{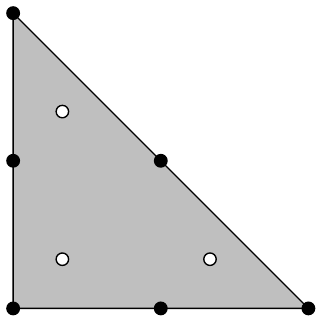}
    & $ \times $
    & \includegraphics{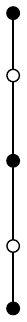}
    & $ = $
    & \includegraphics{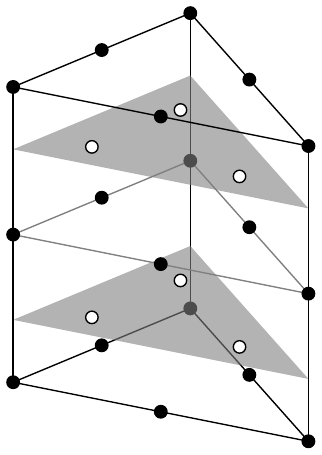}
  \end{tabular}
  \caption{Second order triangular prism element and quadrature
    points.  Black balls denote point evaluation nodes, and white
    balls mark quadrature points.  The element and the quadrature rule
    are both tensor products of triangular and interval elements and
    quadrature rules respectively.  Consequently, the tabulation
    matrix of the prism element is the Kronecker product of the
    tabulation matrices of the triangular and interval elements.}
  \label{fig:wedge}
\end{figure}

When both the finite element and the point set are tensor products of
finite elements and point sets with matching dimensions, as in
\cref{fig:wedge}, FInAT constructs the tabulation expression as the
tensor product of the tabulations of factor elements:
\begin{equation}
  \label{eq:product_subst}
  \Psi_{(i_1,i_2)}(\xi_{(q_1,q_2)}) \mapsto
  \mathcal{E}\left\{\Psi^{(1)}_{i_1}(\xi^{(1)}_{q_1})\right\} \,
  \mathcal{E}\left\{\Psi^{(2)}_{i_2}(\xi^{(2)}_{q_2})\right\} .
\end{equation}
If the factor elements have no further structure, this may simply mean
$ \mathbf{\Psi}^{(1)}_{i_1 q_1} \mathbf{\Psi}^{(2)}_{i_2 q_2} $, where
$ \mathbf{\Psi}^{(1)} $ and $ \mathbf{\Psi}^{(2)} $ are FIAT-provided
tabulation matrices.

\subsection{Collocated quadrature points}
\label{sec:colocated}

Spectral elements \cite{karniadakis2013spectral,Patera1984} use
Lagrange polynomials whose nodes are collocated with quadrature
points.  Typically, Gauss--Lobatto--Legendre (GLL) quadrature points
are chosen for (continuous) interval elements, while Gauss--Legendre
(GL) quadrature points are better for discontinuous elements.  When
these nodes match the quadrature rule that is used to approximate the
integral, then the tabulation matrix becomes the identity matrix.  In
other words, FInAT can return
\begin{equation}
  \label{eq:colocated}
  \Psi_i (\xi_q) \mapsto \delta_{i q},
\end{equation}
and then the form compiler can optimise this away with a resulting
simplification of the loop nests.

Another advantage of spectral elements is that they result in a better
condition number of the assembled linear system than equidistant
Lagrange elements, especially for higher polynomial degrees.
We have added \texttt{GaussLobattoLegendre} and \texttt{GaussLegendre}
elements to FIAT, and the appropriate element wrappers in FInAT.
These wrappers are similar to the generic wrapper in
\cref{sec:fiat_wrapper}, except that they symbolically recognise if
the tabulation points match the element nodes, in which case they
replace the zeroth derivative according to \cref{eq:colocated}.
GLL and GL are defined as one-dimensional quadrature rules, and they
do not directly generalise to simplices.  However, using tensor product
elements one can construct higher-dimensional equivalents on box cells
such as quadrilaterals and hexahedra.

\subsection{Enriched element}
\label{sec:enriched}

By \emph{enriched element} we mean the direct sum two (or more) finite
elements. A basis for such an element is given by the concatenation of the
bases of the summands. Let $ V_1 $ and $ V_2 $
be finite elements on the reference cell $ K $, and let
$ \{ \Psi^{(1)}_{i_1} \}_{i_1=1}^{N_{f,1}} $ and
$ \{ \Psi^{(2)}_{i_2} \}_{i_2=1}^{N_{f,2}} $ be their bases
respectively.  The enriched element $ V := V_1 \oplus V_2 $ has basis
$ \{ \Psi_i \}_{i=1}^{N_{f,1} + N_{f,2}} $ such that
\begin{equation}
  \label{eq:enriched_basis}
  \Psi_i =
  \begin{cases}
    \Psi^{(1)}_i & \quad \mbox{when } 1 \le i \le N_{f,1} , \\
    \Psi^{(2)}_{i - N_{f,1}} & \quad \mbox{when } N_{f,1} < i \le N_{f,1} + N_{f,2} .
  \end{cases}
\end{equation}
For example, the triangular Mini element $ P_2 \oplus B_3 $ denotes
the space of quadratic polynomials enriched by a cubic ``bubble''
function \cite{Arnold1984}.

To implement this element, we utilise the newly introduced
\texttt{Concatenate} node of GEM, which flattens the shape of its
operands and concatenates them.  For example, let $ a $ be a
$ 2 \times 2 $ matrix, $ b $ a scalar, and $ c $ a vector of length 3.
Then $ \mathtt{Concatenate}(a, b, c) $ is a vector of length 8 such
that its first four entries correspond to the entries of $ a $, its
fifth entry to $ b $, and its last three entries to $ c $.  We use
this node to concatenate the basis functions of the subelements.

Let $ \mathcal{E}^{(1)}_{i_1 q} $ and $ \mathcal{E}^{(2)}_{i_2 q} $
denote the tabulation expressions for subelements $ V_1 $ and $ V_2 $.  To
correctly use \texttt{Concatenate}, basis function indices must become
\emph{shape} first, so the tabulation of $ V $ is
\begin{equation}
  \label{eq:enriched_subst_scalar}
  {\mathtt{Concatenate} \left(
      {\big] \mathcal{E}^{(1)}_{i_1 q} \big[}_{i_1} ,
      {\big] \mathcal{E}^{(2)}_{i_2 q} \big[}_{i_2}
    \right)}_i .
\end{equation}
In the general case, any of $ q $, $ i_1 $, and $ i_2 $ can be a
multi-index.  Moreover, enriched elements are not limited to
scalar-valued finite elements, however, all subelements must have the same
value shape.  With value multi-index $ \kappa $,
\cref{eq:enriched_subst_scalar} generalises to
\begin{equation}
  \label{eq:enriched_subst_tensor}
  {\mathtt{Concatenate} \left(
      {\big] \mathcal{E}^{(1)}_{i_1 q \kappa} \big[}_{i_1} ,
      {\big] \mathcal{E}^{(2)}_{i_2 q \kappa} \big[}_{i_2}
    \right)}_i .
\end{equation}
Note that $ q $ and $ \kappa $ just pass through as free
multi-indices, and only the basis function multi-indices $ i_1 $ and
$ i_2 $ are replaced by the unified index $ i $.  We later show
how enriched elements are destructured in the form compiler to recover
the structure within the subelements.

\subsection{Value modifier element wrappers}
\label{sec:hdivcurl}

The facilities of FInAT for tensor product and enriched elements allow
us to implement \Hdiv{} and \Hcurl{} elements in a structure-revealing
way.  These elements are important in
stable mixed finite element discretisations.  Firedrake supports the
whole $ \mathcal{Q}^- $ family of the \emph{Periodic Table of the
  Finite Elements}~\cite{Arnold2014}.  This family is defined on cube
cells, and its members can be constructed out of interval elements.
The continuous Q and the discontinuous dQ elements are \emph{just}
tensor products of interval elements, while the construction of \Hdiv{} and
\Hcurl{} conforming elements is slightly more complicated.
\citet{McRae2014} provide constructions for \Hdiv{} and \Hcurl{}
elements. For example quadilateral Raviart-Thomas~\citeyear{Raviart1977} (RTCF) elements can be constructed as
\begin{equation}
  \label{eq:rtcf_impl}
  \mathrm{RTCF}_n =
  \mathtt{HDiv}(\mathrm{P}_n \otimes \mathrm{dP}_{n-1}) \oplus
  \mathtt{HDiv}(\mathrm{dP}_{n-1} \otimes \mathrm{P}_n)
\end{equation}
where $ \otimes $ denotes tensor product, $ \oplus $ implies an
enriched element, and $ \mathtt{HDiv} $ is a special element wrapper
that applies a transformation to the values of the basis functions of
a tensor product element.
Concretely, if $ \Psi_i(\hat{x}) $ is a basis function of
$ \mathrm{P}_n \otimes \mathrm{dP}_{n-1} $, then the corresponding
basis function of
$ \mathtt{HDiv}(\mathrm{P}_n \otimes \mathrm{dP}_{n-1}) $ is
\begin{equation}
  \label{eq:hdiv_01}
  \Psi^*_i(\hat{x}) =
  \begin{bmatrix}
    -\Psi_i(\hat{x}) \\ 0
  \end{bmatrix}.
\end{equation}
Note that $ \Psi_i $ is a scalar-valued function, while $ \Psi^*_i $
is a vector field.  If $ \mathtt{HDiv} $ is applied to
$ \mathrm{dP}_{n-1} \otimes \mathrm{P}_n $, then
\begin{equation}
  \label{eq:hdiv_10}
  \Psi^*_i(\hat{x}) =
  \begin{bmatrix}
    0 \\ \Psi_i(\hat{x})
  \end{bmatrix}.
\end{equation}
This difference may seem surprising, but it is possible because
$ \mathtt{HDiv} $ is defined as a long switch-case: based on the
continuity and occasionally the reference value transformation of each
factor element, it applies the transformation that makes the result
suitable for building an \Hdiv{} conforming element.
$ \mathtt{HCurl} $ behaves similarly for building \Hcurl{} conforming
elements.

We have listed only two cases of $ \mathtt{HDiv} $ in
\cref{eq:hdiv_01,eq:hdiv_10}.  For all cases of $ \mathtt{HDiv} $ and
$ \mathtt{HCurl} $, as well as for a complete description of the
construction
of $ \mathcal{Q}^- $ family elements, we refer to \citet{McRae2014}.
What is important at the moment, is that these element wrappers
\begin{itemize}
\item do not change or destroy the structure of the element they are
  applied to.  (In the general case, $ i $ is a multi-index.)
\item allow us to build the \Hdiv{} and \Hcurl{} conforming elements
  of the $ \mathcal{Q}^- $ family, together with tensor product and
  enriched elements.
\end{itemize}

\section{Form compiler algorithms}
\label{sec:tsfc}

Expressing the inherent structure of finite elements is necessary,
but not sufficient for achieving optimal code generation.  The
other main ingredient is algorithms for rearranging \emph{tensor
  contractions} (called \texttt{IndexSum} in UFL and GEM) such that an
optimal assembly algorithm is achieved.

These algorithms were implemented in the \emph{Two-Stage Form
  Compiler} (TSFC) \cite{Homolya2017},
the form compiler used in Firedrake.  In the first stage, TSFC
lowers the finite element objects and geometric terms in weak form,
and produces a tensor algebra expression in GEM.  In the second stage,
efficient C code is generated for the evaluation of GEM expression.

TSFC originally used FIAT for an implementation of finite elements, however
changes have been made to the first stage to use FInAT instead.  Since
FInAT provides tabulations as GEM expressions, they integrate
seamlessly into the intermediate representation of TSFC.  The
algorithms that rearrange tensor contractions were implemented as
GEM-to-GEM transformers.  The second stage has been largely left
intact.

\subsection{Delta cancellation: simple case}
\label{sec:delta_simple}

One important application of \emph{delta cancellation}, by which we mean the
simplification of loop nests involving Kronecker deltas, is found in our
handling of mesh coordinates.  The coordinate field is often stored in
numerical software by assigning coordinates to each vertex of the
mesh.  This is equivalent to a
vector-$ P_1 $ or vector-$ Q_1 $ finite element space.  In fact,
Firedrake represents coordinates as an ordinary finite element field,
which facilitates support for higher-order geometries.

Since integral scaling involves the Jacobian of the coordinate
transformation, virtually every finite element kernel is parametrised
by this vector field.  We have seen in \cref{sec:tensor_element} that
the FInAT implementation of vector elements contains Kronecker delta
nodes; we now show how to simplify them away.

Consider a single entry $ J_{a b} $ of the Jacobian matrix.  We need
its evaluation at each quadrature point $ \xi_q $, let us call this
$ j_q := J_{a b}(\xi_q) $.  Let $ \{ \Phi_i \}_{i=1}^{N_c} $ be the
basis functions of the coordinate element, and let
$ \{ c_i \}_{i=1}^{N_c} $ denote the local coefficients of this basis,
so
\begin{equation}
  j_q := J_{a b}(\xi_q) = \sum_{i=1}^{N_c} c_i {\left[ \frac{\partial
        \Phi_i}{\partial \hat{x}_b}(\xi_q) \right]}_{a}.
\end{equation}
Suppose we have a vector-$ P_2 $ coordinate element, then FInAT
gives:
\begin{equation}
  {\left[ \frac{\partial \Phi_i}{\partial \hat{x}_b}(\xi_q) \right]}_a
  \mapsto \mathbf{\Phi}^{(b,*)}_{i_1 q} \delta_{i_2 a}
\end{equation}
where $ i = (i_1, i_2) $ and $ \mathbf{\Phi}^{(b,*)} $ is the
tabulation expression of the $ b $-th derivative of the \emph{scalar}
$ P_2 $ element.  Applying the substitution yields:
\begin{equation}
  j_q = \sum_{i_1,i_2} c_{(i_1,i_2)} \mathbf{\Phi}^{(b,*)}_{i_1 q}
  \delta_{i_2 a} .
\end{equation}
Then the tensor product is disassembled into
\begin{itemize}
\item factors: $ c_{(i_1,i_2)} $,
  $ \mathbf{\Phi}^{(b,*)}_{i_1 q} $, $ \delta_{i_2 a} $; and
\item contraction indices: $ i_1 $, $ i_2 $.
\end{itemize}

Delta cancellation is carried out in this disassembled form.  If there
are any factors $ \delta_{jk} $ or $ \delta_{kj} $ such that $ j $ is
a contraction index, that factor is removed along with the contraction
index $ j $, and a $ j \mapsto k $ index substitution is applied to
all the remaining factors.  This is repeated as long as applicable,
and the repetition trivially terminates at latest when running out of
contraction indices or delta factors.

This cancellation step is only applicable once to the example above,
and it leaves us with factors $ c_{(i_1,a)} $ and
$ \mathbf{\Phi}^{(b,*)}_{i_1 q} $ as well as contraction index
$ i_1 $.  Therefore, the optimised tensor product becomes
\begin{equation}
  j_q = \sum_{i_1} c_{(i_1,a)} \mathbf{\Phi}^{(b,*)}_{i_1 q} .
\end{equation}

As one can see, this only needs the tabulation matrix of the scalar
element, and each entry of the Jacobian only uses the relevant section
of the array $ c $.

\subsection{Sum factorisation: Laplace operator}
\label{sec:sum_fact_poisson}

\emph{Sum factorisation} is a well-known technique first proposed by
\citet{Orszag1980} that drastically reduces the algorithmic
complexity of finite element assembly for high-order discretisations.
To demonstrate this technique, consider the Laplace operator as
introduced in \cref{sec:background}, and let the reference cell
$ \hat{K} $ be a square.  This simplifies our notation, although TSFC
supports higher dimensions uniformly.  During UFL preprocessing, matrix-vector
multiplications and the inner product in \cref{eq:laplace_prefiat} are
rewritten as
\begin{equation}
  A^K_{ij} = \sum_{q=1}^{N_q} w_q S_q \sum_{k=1}^2 \left( \sum_{l=1}^2
    G_{q l k} \frac{\partial \Psi_i}{\partial \hat{x}_l}(\xi_q)
  \right) \left( \sum_{m=1}^2 G_{q m k} \frac{\partial
      \Psi_j}{\partial \hat{x}_m}(\xi_q) \right) .
\end{equation}
To be able to demonstrate a more generic case, we expand the geometric
sums:
\begin{equation}
  \label{eq:poisson_unrolled}
  {\small
  \begin{aligned}
  A^K_{ij} = \sum_{q=1}^{N_q} w_q S_q \left\{
    \left(
      G_{q 1 1} \frac{\partial \Psi_i}{\partial \hat{x}_1}(\xi_q) +
      G_{q 2 1} \frac{\partial \Psi_i}{\partial \hat{x}_2}(\xi_q)
    \right) \left(
      G_{q 1 1} \frac{\partial \Psi_j}{\partial \hat{x}_1}(\xi_q) +
      G_{q 2 1} \frac{\partial \Psi_j}{\partial \hat{x}_2}(\xi_q)
    \right)
  \right. \\ \left.
    +
    \left(
      G_{q 1 2} \frac{\partial \Psi_i}{\partial \hat{x}_1}(\xi_q) +
      G_{q 2 2} \frac{\partial \Psi_i}{\partial \hat{x}_2}(\xi_q)
    \right) \left(
      G_{q 1 2} \frac{\partial \Psi_j}{\partial \hat{x}_1}(\xi_q) +
      G_{q 2 2} \frac{\partial \Psi_j}{\partial \hat{x}_2}(\xi_q)
    \right)
  \right\} .
  \end{aligned}
  }
\end{equation}

With a tensor product element and quadrature rule
as introduced in \cref{eq:tpps,eq:limits_tpe}, FInAT provides the
substitutions
\begin{align}
  \frac{\partial \Psi_i}{\partial \hat{x}_1}(\xi_q)
  &\mapsto D\mathbf{\Psi}^{(1)}_{i_1 q_1} \mathbf{\Psi}^{(2)}_{i_2 q_2} \quad \text{and} \\
  \frac{\partial \Psi_i}{\partial \hat{x}_2}(\xi_q)
  &\mapsto \mathbf{\Psi}^{(1)}_{i_1 q_1} D\mathbf{\Psi}^{(2)}_{i_2 q_2}
\end{align}
where $ i = (i_1, i_2) $, and $ \mathbf{\Psi}^{(1)} $ and
$ \mathbf{\Psi}^{(2)} $ are numerical tabulation matrices of the
one-dimensional elements, while $ D\mathbf{\Psi}^{(1)} $ and
$ D\mathbf{\Psi}^{(2)} $ are tabulations of the first derivative
respectively.  Thus \cref{eq:poisson_unrolled} becomes
\begin{equation}
  \label{eq:poisson_gem}
  \begin{aligned}
  A^K_{(i_1,i_2) (j_1,j_2)} =
    \textstyle{\sum}_{q_1,q_2} w^{(1)}_{q_1} w^{(2)}_{q_2} S_{q_1 q_2} & \left\{
      \left(
        G_{q_1 q_2 1 1} D\mathbf{\Psi}^{(1)}_{i_1 q_1} \mathbf{\Psi}^{(2)}_{i_2 q_2} +
        G_{q_1 q_2 2 1} \mathbf{\Psi}^{(1)}_{i_1 q_1} D\mathbf{\Psi}^{(2)}_{i_2 q_2}
        \right) \right. \\
        & \quad \left(
      G_{q_1 q_2 1 1} D\mathbf{\Psi}^{(1)}_{j_1 q_1} \mathbf{\Psi}^{(2)}_{j_2 q_2} +
      G_{q_1 q_2 2 1} \mathbf{\Psi}^{(1)}_{j_1 q_1} D\mathbf{\Psi}^{(2)}_{j_2 q_2}
    \right)
    \\ & + \left(
      G_{q_1 q_2 1 2} D\mathbf{\Psi}^{(1)}_{i_1 q_1} \mathbf{\Psi}^{(2)}_{i_2 q_2} +
      G_{q_1 q_2 2 2} \mathbf{\Psi}^{(1)}_{i_1 q_1} D\mathbf{\Psi}^{(2)}_{i_2 q_2}
    \right) \\ & \quad \left. \left(
        G_{q_1 q_2 1 2} D\mathbf{\Psi}^{(1)}_{j_1 q_1} \mathbf{\Psi}^{(2)}_{j_2 q_2} +
        G_{q_1 q_2 2 2} \mathbf{\Psi}^{(1)}_{j_1 q_1} D\mathbf{\Psi}^{(2)}_{j_2 q_2}
      \right)
    \right\} .
  \end{aligned}
\end{equation}


Now we reached the intermediate representation of TSFC:
all finite element basis functions have been replaced with tensor
algebra expressions.  Before we can apply sum factorisation, 
we need to apply \emph{argument factorisation} to
\cref{eq:poisson_gem}.  Argument factorisation transforms the expression to a
sum-of-products form, such that no factor in any product depends on
more than one of the free indices of form expression.  In the above
example, the free indices are $ i_1 $, $ i_2 $, $ j_1 $, and $ j_2 $.
\Cref{eq:poisson_gem} after argument factorisation looks like
\begin{equation}
  \begin{aligned}
    A^K_{(i_1,i_2) (j_1,j_2)} = \textstyle{\sum}_{q_1,q_2}
    & \left\{
      D\mathbf{\Psi}^{(1)}_{i_1 q_1} \mathbf{\Psi}^{(2)}_{i_2 q_2}
      D\mathbf{\Psi}^{(1)}_{j_1 q_1} \mathbf{\Psi}^{(2)}_{j_2 q_2}
      P^{(1,1)}_{q_1 q_2}
    \right. \\
    & \quad +
    D\mathbf{\Psi}^{(1)}_{i_1 q_1} \mathbf{\Psi}^{(2)}_{i_2 q_2}
    \mathbf{\Psi}^{(1)}_{j_1 q_1} D\mathbf{\Psi}^{(2)}_{j_2 q_2}
    P^{(1,2)}_{q_1 q_2} \\
    & \quad +
    \mathbf{\Psi}^{(1)}_{i_1 q_1} D\mathbf{\Psi}^{(2)}_{i_2 q_2}
    D\mathbf{\Psi}^{(1)}_{j_1 q_1} \mathbf{\Psi}^{(2)}_{j_2 q_2}
    P^{(2,1)}_{q_1 q_2} \\
    & \quad \left. +
      \mathbf{\Psi}^{(1)}_{i_1 q_1} D\mathbf{\Psi}^{(2)}_{i_2 q_2}
      \mathbf{\Psi}^{(1)}_{j_1 q_1} D\mathbf{\Psi}^{(2)}_{j_2 q_2}
      P^{(2,2)}_{q_1 q_2}
    \right\}
  \end{aligned}
\end{equation}
where $ P^{(1,1)}_{q_1 q_2} $, $ P^{(1,2)}_{q_1 q_2} $,
$ P^{(2,1)}_{q_1 q_2} $, and $ P^{(2,2)}_{q_1 q_2} $ aggregate factors
that do not depend on any of the free indices.  Their precise
definitions are:
\begin{subequations}
  \begin{align}
    P^{(1,1)}_{q_1 q_2} &:= w^{(1)}_{q_1} w^{(2)}_{q_2} S_{q_1 q_2}
                          ( G_{q_1 q_2 1 1} G_{q_1 q_2 1 1} +
                          G_{q_1 q_2 1 2} G_{q_1 q_2 1 2} ) \\
    P^{(1,2)}_{q_1 q_2} &:= w^{(1)}_{q_1} w^{(2)}_{q_2} S_{q_1 q_2}
                          ( G_{q_1 q_2 1 1} G_{q_1 q_2 1 2} +
                          G_{q_1 q_2 1 2} G_{q_1 q_2 2 2} ) \\
    P^{(2,1)}_{q_1 q_2} &:= w^{(1)}_{q_1} w^{(2)}_{q_2} S_{q_1 q_2}
                          ( G_{q_1 q_2 1 1} G_{q_1 q_2 1 2} +
                          G_{q_1 q_2 1 2} G_{q_1 q_2 2 2} ) \\
    P^{(2,2)}_{q_1 q_2} &:= w^{(1)}_{q_1} w^{(2)}_{q_2} S_{q_1 q_2}
                          ( G_{q_1 q_2 2 1} G_{q_1 q_2 2 1} +
                          G_{q_1 q_2 2 2} G_{q_1 q_2 2 2} )
  \end{align}
\end{subequations}

Were there no restrictions on the factors, any expression would
trivially be in a sum-of-products form.  One can reach the argument
factorised form through the mechanical application of the
distributive property $ a (b + c) \rightarrow a b + a c $ on products
which have any factor with more than one free index.  This rewriting
always succeeds since the variational form is by definition linear in
its arguments.

To continue with sum factorisation, let us concentrate on just one of
the products for now.  For example:
\begin{equation}
  \label{eq:single_product}
  \sum_{q_1,q_2} D\mathbf{\Psi}^{(1)}_{i_1 q_1}
  \mathbf{\Psi}^{(2)}_{i_2 q_2} D\mathbf{\Psi}^{(1)}_{j_1 q_1}
  \mathbf{\Psi}^{(2)}_{j_2 q_2} P^{(1,1)}_{q_1 q_2}
\end{equation}
This requires $ O(N_{q_1} N_{q_2} N_{f_1} N_{f_2} N_{f_1} N_{f_2}) $
floating-point operations to evaluate.  If the finite element is of
polynomial order $ n $ in both directions, then
$ N_{f_1} = N_{f_2} = n + 1 $, and also $ N_{q_1} = N_{q_2} = O(n) $.
This means $ O(n^6) $ operations.
Rearranging \cref{eq:single_product} as
\begin{equation}
  \label{eq:opt1_product}
  \sum_{q_1=1}^{N_{q_2}} D\mathbf{\Psi}^{(1)}_{i_1 q_1}
  D\mathbf{\Psi}^{(1)}_{j_1 q_1} \left( \sum_{q_2=1}^{N_{q_2}}
    \mathbf{\Psi}^{(2)}_{i_2 q_2} \mathbf{\Psi}^{(2)}_{j_2 q_2}
    P^{(1,1)}_{q_1 q_2} \right)
\end{equation}
only requires
$ O(N_{q_1} N_{q_2} N^2_{f_2} + N_{q_1} N^2_{f_1} N^2_{f_2}) $
operations, and rearranging as
\begin{equation}
  \label{eq:opt2_product}
  \sum_{q_2=1}^{N_{q_2}} \mathbf{\Psi}^{(2)}_{i_2 q_2}
  \mathbf{\Psi}^{(2)}_{j_2 q_2} \left( \sum_{q_1=1}^{N_{q_2}}
    D\mathbf{\Psi}^{(1)}_{i_1 q_1} D\mathbf{\Psi}^{(1)}_{j_1 q_1}
    P^{(1,1)}_{q_1 q_2} \right)
\end{equation}
only requires
$ O(N_{q_1} N_{q_2} N^2_{f_1} + N_{q_2} N^2_{f_1} N^2_{f_2}) $
operations.  Both arrangements imply $ O(n^5) $ operations.  Since the
number of products in the argument factorised form only depends on the
original weak form, but not on the polynomial order $ n $, it is easy
to see that assembling a bilinear form on a single quadrilateral
requires $ O(n^6) $ operations na{\"i}vely, or $ O(n^5) $ operations
with sum factorisation.  Generally for a $ d $-cube the operation
count is $ O(n^{3 d}) $ na{\"i}vely, and $ O(n^{2 d + 1}) $ with sum
factorisation.

Finally, the only missing piece is a systematic algorithm that takes a
tensor product such as \cref{eq:single_product}, and rearranges it
into an optimal tensor product such as either \cref{eq:opt1_product}
or \cref{eq:opt2_product}.  We first disassemble the tensor product
into factors and contraction indices as in \cref{sec:delta_simple},
and then build an optimised tensor product from them.  This problem
has also been relevant in quantum chemistry applications, and is known
as \emph{single-term optimisation} in the literature.  \citet{Lam1996}
prove that this problem is NP-complete.  Their algorithm explores
combinations of factors to build a product tree, applying contractions
on the way, and pruning the search tree to avoid traversing redundant
expressions.  Our approach, described in \cref{alg:sum_fact}, orders
the contraction indices first: since we never have more than three of
them, traversing all permutations is cheap.

\begin{algorithm}
  \begin{algorithmic}[1]
    \Function{MakeTensorProduct}{$ \mathtt{factors} $, $ \mathtt{indices} $}
    \State $ \mathtt{result} \leftarrow \mathit{nil} $
    \State $ \mathtt{minops} \leftarrow \infty $
    \ForAll{$ \mathtt{ordering} \in \textsc{Permutations}(\mathtt{indices}) $}
      \State $ \mathtt{terms} \leftarrow \mathtt{factors} $ \label{lin:perm_start}
      \State $ \mathtt{flops} \leftarrow 0 $
      \For{$ \mathtt{index} $ in $ \mathtt{ordering} $}
      \Comment{apply contraction, one index at a time}
      \State $ \mathtt{contract} \leftarrow \{ t \in \mathtt{terms}
      \mid \mathtt{index} \in \text{free indices of } t \} $ \label{lin:contract}
      \State $ \mathtt{deferred} \leftarrow \{ t \in \mathtt{terms}
      \mid \mathtt{index} \notin \text{free indices of } t \} $ \label{lin:deferred}
      \Statex
      \State $ \mathtt{product}, \mathtt{cost} \leftarrow
      \textsc{MakeProduct}(\mathtt{contract}) $
      \State $ \mathtt{term} \leftarrow $ \texttt{IndexSum(product, (index,))}
      \Statex
      \State $ \mathtt{terms} \leftarrow \mathtt{deferred} \cup \{
      \mathtt{term} \} $ \label{lin:update_terms}
      \State $ \mathtt{flops} \leftarrow \mathtt{flops} +
      \mathtt{cost} + \prod_{i \in \text{free indices of }
        \mathtt{product}} i\mathtt{.extent} $
    \EndFor
    \State $ \mathtt{expr}, \mathtt{cost} \leftarrow
    \textsc{MakeProduct}(\mathtt{terms}) $ \label{lin:multiple_remain_start}
    \State $ \mathtt{flops} \leftarrow \mathtt{flops} + \mathtt{cost} $
    \label{lin:multiple_remain_end} \label{lin:perm_end}
    \Statex
    \If{$ \mathtt{flops} < \mathtt{minops} $}
      \State $ \mathtt{result} \leftarrow \mathtt{expr} $
      \State $ \mathtt{minops} \leftarrow \mathtt{flops} $
    \EndIf
    \EndFor
    \State \Return $ \mathtt{result} $
    \EndFunction
  \end{algorithmic}
  \caption{Building an optimal tensor product}
  \label{alg:sum_fact}
\end{algorithm}

The construction of an optimised tensor product expression happens in
\crefrange{lin:perm_start}{lin:perm_end} of \cref{alg:sum_fact}.  We
apply contractions one index at a time.  For example, let us contract
along $ q_1 $ first.  Then the set of factors are split based on
dependence on $ q_1 $ (\cref{lin:contract,lin:deferred}).  The
factors in \texttt{deferred} are ``pulled out'' of
$ \sum_{q_1} $, while a product expression is constructed from
the factors in \texttt{contract} using \textsc{MakeProduct}, followed
by contraction over $ q_1 $.  \textsc{MakeProduct} is a utility
function for building a product expression tree from a set of factors,
also returning the number of multiplications required to evaluate that
product.  For further gains, this function may associate factors in a
optimised way, but for sum factorisation a trivial product builder
suffices. For our example
\begin{align}
  \mathtt{deferred} &= \{ \mathbf{\Psi}^{(2)}_{i_2 q_2} ,
                      \mathbf{\Psi}^{(2)}_{j_2 q_2} \} \qquad \text{and} \\
  \mathtt{term} &= \sum_{q_1} D\mathbf{\Psi}^{(1)}_{i_1 q_1}
                  D\mathbf{\Psi}^{(1)}_{j_1 q_1} P^{(1,1)}_{q_1 q_2} .
\end{align}
The new tensor contraction expression (\texttt{term}) is added to the
deferred factors to produce the set of factors (\texttt{terms}) for
the next iteration (\cref{lin:update_terms}).  In the next iteration
we contract over $ q_2 $ and finally get \cref{eq:opt2_product}.
For generality, we also handle the case in
\crefrange{lin:multiple_remain_start}{lin:multiple_remain_end} when
multiple factors remain after applying all contraction indices.  This
construction is repeated for each permutation of the contraction
indices, and the expression requiring the lowest number of
floating-point operations is selected.  Looping over all permutations
is not that bad, since $ \left| \mathtt{indices} \right| \le 3 $ for
all relevant problems.

\subsection{Sum factorisation: parametrised forms}
\label{sec:sum_fact_coeff}

Partial differential equations are frequently parametrised by
prescribed spatial functions, which are called \emph{coefficients} in
UFL.  We have shown that matrix-free assembly of the operator action
is analogous to assembling
forms with coefficient functions, so as an example we now consider the
action of the Laplace operator.  We continue the description in
\cref{sec:background} from \cref{eq:laplace_quadrature}, but now
\begin{equation}
  \label{eq:coeff_def}
  \hat{u}_K(\hat{x}) = \sum_{i=1}^{N_f} U^K_i \Psi_i(\hat{x})
\end{equation}
where $ U^K \in \mathbb{R}^{N_f} $ is an array of local basis function
coefficients.  This results in
\begin{equation}
  b^K_{j} = \sum_{q=1}^{N_q} w_q S_q \left( G_q^T \sum_{i=1}^{N_f}
    U^K_i \hat{\nabla} \Psi_i(\xi_q) \right) \cdot \left( G_q^T
    \hat{\nabla} \Psi_j(\xi_q) \right)
\end{equation}
where $ b^K $ is the assembled local vector.

The common strategy is to first evaluate the coefficient function as
$ C_q \in \mathbb{R}^d $ at every quadrature point $ q $, then simply
use that in the form expression:
\begin{align}
  \label{eq:coeff_eval}
  C_q &= \sum_{i=1}^{N_f} U^K_i \hat{\nabla} \Psi_i(\xi_q) \\
  \label{eq:form_eval}
  b^K_j &= \sum_{q=1}^{N_q} w_q S_q ( G_q^T C_q ) \cdot ( G_q^T
          \hat{\nabla} \Psi_j(\xi_q) )
\end{align}
The sum factorisation of \cref{eq:form_eval} is basically analogous to
that which was shown in \cref{sec:sum_fact_poisson}.  \Cref{eq:coeff_eval}
is even simpler: it is similar to the products of the argument
factorised form, so it just needs disassembling and calling
\cref{alg:sum_fact}.

\Cref{tab:form_ordo} summarises the gains of sum factorisation
considering various cell types, both for bilinear forms (matrix
assembly) as well as for linear forms such as matrix-free operator
actions or right-hand side assembly.

\begin{table}[tbhp]
  \captionsetup{position=top}
  \caption{Algorithmic complexity of form assembly on a single cell as
    a function of polynomial degree $ n $.}
  \label{tab:form_ordo}
  \vspace{\baselineskip}
  \centering
  \subfloat[Na{\"i}ve implementation]{
    \label{tab:form_ordo_naive}
    \begin{tabular}{l | c | c}
      Cell type & \textbf{\small Linear form} & \textbf{\small Bilinear form} \\
      \hline
      {\small quadrilateral} & $ O(n^4) $ & $ O(n^6) $ \\
      {\small triangular prism} & $ O(n^6) $ & $ O(n^9) $ \\
      {\small hexahedron} & $ O(n^6) $ & $ O(n^9) $
    \end{tabular}
  }
  \enspace
  \subfloat[Sum factorised implementation]{
    \label{tab:form_ordo_sumfac}
    \begin{tabular}{l | c | c}
      Cell type & \textbf{\small Linear form} & \textbf{\small Bilinear form} \\
      \hline
      {\small quadrilateral} & $ O(n^3) $ & $ O(n^5) $ \\
      {\small triangular prism} & $ O(n^5) $ & $ O(n^7) $ \\
      {\small hexahedron} & $ O(n^4) $ & $ O(n^7) $
    \end{tabular}
  }
\end{table}

\subsection{Delta cancellation: nonzero patterns}
\label{sec:delta_nonzero}

In \cref{sec:delta_simple}, we considered delta cancellation for a
common, but simple case; now we explore delta cancellation further.
Previously, we could assume that the assembled local tensors were
dense; now we are going to see that \emph{Kronecker delta} nodes may
cause particular nonzero patterns to appear.

For a concrete example, consider the vector mass form
\begin{equation}
  \label{eq:vector_mass_form}
  \int_\Omega u \cdot v \dx
\end{equation}
where the trial function $ u $ and the test function $ v $ are chosen
from vector-$ P_n $ elements.  Going through the usual steps, the
integral on cell $ K $ is evaluated as
\begin{equation}
  \label{eq:vector_mass_prefiat}
  A^K_{ij} = \sum_{q=1}^{N_q} w_q S_q \Psi_i(\xi_q) \cdot
  \Psi_j(\xi_q) .
\end{equation}
Recall that the usual translation of vector-$ P_n $ elements is
\begin{equation}
  {\left[ \Psi_i(\xi_q) \right]}_k \mapsto \mathbf{\Psi}^*_{i_1 q} \delta_{i_2 k}
\end{equation}
where $ i = (i_1, i_2) $ and $ \mathbf{\Psi}^* $ is the tabulation
matrix of the \emph{scalar} $ P_n $ element.  Applying the
substitution and rewriting the dot product we have
\begin{equation}
  \label{eq:vector_mass_gem}
  A^K_{(i_1,i_2) (j_1,j_2)} = \sum_{q=1}^{N_q} w_q S_q \sum_{k=1}^d
  \mathbf{\Psi}^*_{i_1 q} \delta_{i_2 k} \mathbf{\Psi}^*_{j_1 q}
  \delta_{j_2 k} .
\end{equation}
If we apply delta cancellation as described in \cref{sec:delta_simple}
to this product, we obtain
\begin{equation}
  A^K_{(i_1,i_2) (j_1,j_2)} = \sum_{q=1}^{N_q} w_q S_q
  \mathbf{\Psi}^*_{i_1 q} \mathbf{\Psi}^*_{j_1 q} \delta_{i_2 j_2} ,
\end{equation}
or if we build the tensor product with sum factorisation, even
\begin{equation}
  A^K_{(i_1,i_2) (j_1,j_2)} = \delta_{i_2 j_2} \sum_{q=1}^{N_q} w_q
  S_q \mathbf{\Psi}^*_{i_1 q} \mathbf{\Psi}^*_{j_1 q} .
\end{equation}

It is clear that $ A^K $ has a particular nonzero pattern, that is
$ A^K_{(i_1,i_2) (j_1,j_2)} \ne 0 $ only if $ i_2 = j_2 $.  To exploit
this property, we need another delta cancellation step that operates
\emph{across the assignment}.  That is, the tensor product is
disassembled and deltas cancelled with contraction indices as in
\cref{sec:delta_simple}.  Then, still in the disassembled form, if
there are any factors $ \delta_{jk} $ or $ \delta_{kj} $ such that
$ j $ is a free index of the \emph{return variable}, then that factor
is removed and a $ j \mapsto k $ index substitution is applied to all
the remaining factors as well as the return variable.  Again, this is
repeated as long as applicable.  Finally, the tensor product is
rebuilt.

In the example above, the return variable is
$ A^K_{(i_1,i_2) (j_1,j_2)} $, and its free indices are $ i_1 $,
$ i_2 $, $ j_1 $, and $ j_2 $.  Cancelling the remaining
$ \delta_{i_2 j_2} $, we end up with
\begin{equation}
  A^K_{(i_1,i_2) (j_1,i_2)} = \sum_{q=1}^{N_q} w_q S_q
  \mathbf{\Psi}^*_{i_1 q} \mathbf{\Psi}^*_{j_1 q} .
\end{equation}
Note the change in indexing the left-hand side.

In the general case, however, we will not have a nice product
structure as in \cref{eq:vector_mass_gem}.  Nevertheless, if we apply
argument factorisation as in \cref{sec:sum_fact_poisson}, then delta
cancellation as described above can be applied to each product.  Note
that applying delta cancellation across assignments may cause
different parts of the form expression to be ``assigned'' to different
\emph{views} of the return variable.  A simple approach to correctly
handle this case is to assume that the buffer holding $ A^K $ is
cleared at the beginning, and then make each ``assignment'' add to
that buffer.

\subsection{Splitting \texttt{Concatenate} nodes}
\label{sec:split_concat}

\newcommand{\RtcfTwoX}[0]{\mathtt{HDiv}(\mathrm{P}_2 \otimes \mathrm{dP}_1)}
\newcommand{\RtcfTwoY}[0]{\mathtt{HDiv}(\mathrm{dP}_1 \otimes \mathrm{P}_2)}

\begin{figure}
  \centering
  \begin{tikzpicture}[scale=2.5]
    \draw (0,0) rectangle (1,1);
    \pgfmathsetmacro{\c}{1}
    \foreach \x in {0.0, 0.5, 1.0} {
      \foreach \y in {0.211, 0.789} {
        \filldraw (\x,\y) circle (0.6pt) node (n) {};
        \draw[->,>=stealth] (n) -- ++(-5pt,0);
        \node[above left] (l) at (n) {\small \textbf{\c}};
        \pgfmathsetmacro{\cc}{int(\c + 1)}
        \global\let\c=\cc
      }
    }
    \foreach \x in {0.211, 0.789} {
      \foreach \y in {0.0, 0.5, 1.0} {
        \filldraw (\x,\y) circle (0.6pt) node (n) {};
        \draw[->,>=stealth] (n) -- ++(0,5pt);
        \node[above left] (l) at (n) {\small \textbf{\c}};
        \pgfmathsetmacro{\cc}{int(\c + 1)}
        \global\let\c=\cc
      }
    }
    \draw[dotted]
    (0, 0.5) -- (1, 0.5)
    (0.5, 0) -- (0.5, 1)
    (0.211, 0) -- (0.211, 1)
    (0.789, 0) -- (0.789, 1)
    (0, 0.211) -- (1, 0.211)
    (0, 0.789) -- (1, 0.789);
  \end{tikzpicture}
  \caption{Degrees of freedom of the $ \mathrm{RTCF}_2 $ element.}
  \label{fig:rtcf2}
\end{figure}
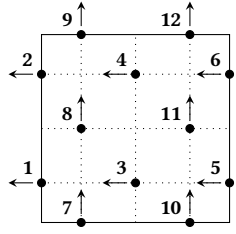

Finally, we consider \texttt{Concatenate} nodes which come from
enriched elements, and are crucial for the \Hdiv{} and \Hcurl{}
conforming elements of the $ \mathcal{Q}^- $ family.  For the sake of
this discussion, we use the $ \mathrm{RTCF}_2 $ element as an example.
As a corollary of \cref{eq:rtcf_impl}, this element is built as
\begin{equation}
  \label{eq:rtcf2}
  \mathrm{RTCF}_2 = \RtcfTwoX \oplus \RtcfTwoY .
\end{equation}
Its degrees of freedom consist of a $ 3 \times 2 $ matrix for
$ \mathrm{P}_2 \otimes \mathrm{dP}_1 $, and a $ 2 \times 3 $ matrix
for $ \mathrm{dP}_1 \otimes \mathrm{P}_2 $; a total of 12 degrees of
freedom.  This is graphically shown in \cref{fig:rtcf2}.

Suppose the quadrature rule is a tensor product rule with the same
interval rule in both directions, then FInAT gives the tabulation
expression:
\begin{equation}
  \label{eq:concat_rtcf}
  {\mathtt{Concatenate} \left(
      {\left]
          {(-\mathbf{\Psi}_{i_1 q_1} \mathbf{\Phi}_{i_2 q_2}, 0)}_c
        \right[}_{(i_1,i_2)} ,
      {\left]
          {(0, \mathbf{\Phi}_{i_3 q_1} \mathbf{\Psi}_{i_4 q_2})}_c
        \right[}_{(i_3,i_4)}
    \right)}_I
\end{equation}
where
\begin{itemize}
\item $ \mathbf{\Psi} $ and $ \mathbf{\Phi} $ are the tabulation
  matrices of the $ \mathrm{P}_2 $ and $ \mathrm{dP}_1 $ interval
  elements respectively.
\item $ q_1 $ and $ q_2 $ are quadrature indices with the same extent.
\item $ i_1 $, $ i_2 $, $ i_3 $, and $ i_4 $ are basis function
  indices of the subelements. \\
  Their extents are 3, 2, 2, and 3 respectively.
\item $ c $ is the value index.  ($ \mathrm{RTCF}_2 $ element is
  vector valued.)
\item $ I $ is the flat basis function index, with extent 12.
\end{itemize}
Suppose we have a buffer $ b $ for the degrees of freedom, and that
the above GEM expression is associated with the indexed buffer
$ b_I $.  The principal approach to ``implementing''
\texttt{Concatenate} nodes is to split them in combination with the
indexed buffer expression.  Splitting \cref{eq:concat_rtcf} with
$ b_I $ gives the following pairs:
\begin{align}
  \label{eq:rtcf_split_x}
  b_{i_2 + 2 (i_1 - 1)} :
  & \quad {(-\mathbf{\Psi}_{i_1 q_1} \mathbf{\Phi}_{i_2 q_2}, 0)}_c \\
  \label{eq:rtcf_split_y}
  b_{6 + i_4 + 3 (i_3 - 1)} :
  & \quad {(0, \mathbf{\Phi}_{i_3 q_1} \mathbf{\Psi}_{i_4 q_2})}_c
\end{align}
where $ b_{i_2 + 2 (i_1 - 1)} $ indexes the first 6 entries of $ b $
as if it were $ 3 \times 2 $ matrix indexed by $ (i_1, i_2) $, and
$ b_{6 + i_4 + 3 (i_3 - 1)} $ indexes the last 6 entries of $ b $ as
if it were $ 2 \times 3 $ matrix indexed by $ (i_3, i_4) $.
These indexed buffer
expressions are easily generated: inspecting the shapes of the
\texttt{Concatenate} node's operands, we see which segment of $ b $
corresponds to each operand and how that should be ``reshaped''.

\newcommand{\PsiX}[0]{\Psi^{\textsc{x}}}
\newcommand{\PsiY}[0]{\Psi^{\textsc{y}}}

First, we discuss the splitting of \texttt{Concatenate} nodes which
originate from the translation of trial and test functions.  Suppose
we have a bilinear form such as
$ \int_\Omega (\nabla \times u) \cdot (\nabla \times v) \dx $.
Following the transformation to reference space, the assignment to the
local tensor $ A^K $ can be written as
\begin{equation}
  \label{eq:bilinear_flat}
  A^K_{I,J} :=
  \mathcal{F} \left( \Psi_I, \hat{\nabla}\Psi_I,
    \Psi_J, \hat{\nabla}\Psi_J \right)
  \qquad 1 \le I \le N_f \text{ and } 1 \le J \le N_f
\end{equation}
where $ \{ \Psi_I \}_{I=1}^{N_f} $ is the basis of a finite element,
$ I $ and $ J $ are basis function indices, and $ \mathcal{F} $ is a
functional.  Sometimes generality may require second and further
derivatives.  Suppose the $ \mathrm{RTCF}_2 $ element was chosen as
above, so $ N_f = 12 $.  Similarly, let
$ \{ \PsiX_{i_1 i_2} \}_{(i_1,i_2)} $ be the basis of $ \RtcfTwoX $
and $ \{ \PsiY_{i_3 i_4} \}_{(i_3,i_4)} $ be the basis of
$ \RtcfTwoY $.  Note that
\begin{align}
  \label{eq:concat_subst_x}
  \{ \PsiX_{i_1 i_2} \}_{(i_1,i_2)} &= \{ \Psi_I \}_{I=1}^6
                                      \quad \text{and} \\
  \label{eq:concat_subst_y}
  \{ \PsiY_{i_3 i_4} \}_{(i_3,i_4)} &= \{ \Psi_I \}_{I=7}^{12} .
\end{align}
\Cref{eq:bilinear_flat} can be trivially split into four assignments,
for the combinations of ranges $ 1 \le I \le 6 $ and
$ 7 \le I \le 12 $ with $ 1 \le J \le 6 $ and $ 7 \le J \le 12 $.
Then the substitutions of \cref{eq:concat_subst_x,eq:concat_subst_y}
are directly applicable, which gives:
\begin{subequations}
  \begin{align}
    A^K_{i_2 + 2 (i_1 - 1), \, j_2 + 2 (j_1 - 1)}
    &:= \mathcal{F} \left( \PsiX_{i_1 i_2}, \hat{\nabla}\PsiX_{i_1 i_2},
      \PsiX_{j_1 j_2}, \hat{\nabla}\PsiX_{j_1 j_2} \right) \\
    A^K_{i_2 + 2 (i_1 - 1), \, 6 + j_4 + 3 (j_3 - 1)}
    &:= \mathcal{F} \left( \PsiX_{i_1 i_2}, \hat{\nabla}\PsiX_{i_1 i_2},
      \PsiY_{j_3 j_4}, \hat{\nabla}\PsiY_{j_3 j_4} \right) \\
    A^K_{6 + i_4 + 3 (i_3 - 1), \, j_2 + 2 (j_1 - 1)}
    &:= \mathcal{F} \left( \PsiY_{i_3 i_4}, \hat{\nabla}\PsiY_{i_3 i_4},
      \PsiX_{j_1 j_2}, \hat{\nabla}\PsiX_{j_1 j_2} \right) \\
    A^K_{6 + i_4 + 3 (i_3 - 1), \, 6 + j_4 + 3 (j_3 - 1)}
    &:= \mathcal{F} \left( \PsiY_{i_3 i_4}, \hat{\nabla}\PsiY_{i_3 i_4},
      \PsiY_{j_3 j_4}, \hat{\nabla}\PsiY_{j_3 j_4} \right)
  \end{align}
\end{subequations}

To be precise, TSFC does not inspect the construction of finite
elements: this splitting of assignments is carried out at the GEM
level, that is, after the substitution of basis function evaluations,
thus TSFC only ``sees'' the occurrences of \texttt{Concatenate} nodes.
This discussion, however, helps to demonstrate that the substitution
of indexed \texttt{Concatenate} nodes such as \cref{eq:concat_rtcf}
with their split indexed expressions -- like
\cref{eq:rtcf_split_x,eq:rtcf_split_y} -- inside the intermediate form
expression, along with the corresponding changes to indexing the
result buffer, is a valid transformation that eliminates the
\texttt{Concatenate} nodes and recovers any product structure within
the subelements.

Lastly, we consider the evaluation of parametrising functions -- also
known as \emph{coefficients} of the multilinear form -- at quadrature
points.  Recalling \cref{eq:coeff_def}, we can apply a similar
separation of basis functions and substitution, that is
\begin{align}
  \hat{u}_K(\hat{x})
  &= \sum_{i=1}^{12} U^K_i \Psi_i(\hat{x})
    = \sum_{i=1}^{6} U^K_i \Psi_i(\hat{x})
    + \sum_{i=7}^{12} U^K_i \Psi_i(\hat{x}) \\
  &= \sum_{i_1,i_2} U^K_{i_2 + 2 (i_1 - 1)} \PsiX_{i_1 i_2}(\hat{x})
    + \sum_{i_3,i_4} U^K_{6 + i_4 + 3 (i_3 - 1)} \PsiY_{i_3 i_4}(\hat{x}) .
\end{align}
Having recovered the product structure, sum factorisation is now
applicable to each summation separately.  Again, we split
\texttt{Concatenate} nodes along with indexed buffer expressions, but
the latter now corresponds to an array of given numbers rather than
the result buffer.  Although the indexed \texttt{Concatenate} node is
not necessarily outermost in the tabulation expression
since the enriched element may not be outermost, this is not problem
since all compound elements in \cref{sec:finat} are linear in their
subelements.

\subsection{Order of transformations}

In previous subsections we have described a number of algorithms
transforming the intermediate representation in TSFC.  We finally list
all transformations in the order of their application:
\begin{enumerate}[label={(\arabic*)}]
\item Split \texttt{Concatenate} nodes, see \cref{sec:split_concat}.
\item Apply argument factorisation, see \cref{sec:sum_fact_poisson}.
\item Apply delta cancellation with tensor contractions, see
  \cref{sec:delta_simple}.
\item Apply delta cancellation across assignments, see
  \cref{sec:delta_nonzero}.
\item Apply sum factorisation, see \cref{sec:sum_fact_poisson}. \\
  Argument factorisation created a sum-of-products form, sum
  factorisation is applied on each product, but with the same ordering
  of contraction indices for all products, to leave more opportunities
  for factorisation based on the distributivity rule.
\item At each contraction level during sum factorisation: apply the ILP factorisation algorithm
  from COFFEE~\cite{Luporini2016}.  This factorisation is based on the
  distributivity rule, and especially improves bilinear forms (matrix
  assembly).
\end{enumerate}

TSFC has several optimisation \emph{modes}, which share the same
UFL-to-GEM and GEM-to-C stages, but carry out different GEM-to-GEM
transformations in between.  Currently, TSFC offers the following
modes:
\begin{itemize}
\item \texttt{spectral} mode applies all passes listed above.
  (The current default.)
\item \texttt{coffee} mode implements a simplified version of the
  algorithm developed by \citet{Luporini2016}.
  The simplification is that this mode unconditionally argument
  factorises, so it contains passes (1), (2), and (6).
  (The previous default.)
\item \texttt{vanilla} mode aims to do as little as possible, so it
  only applies pass (1) since \texttt{Concatenate} nodes must be
  removed before the GEM-to-C stage.  This mode corresponds to the
  original behaviour of TSFC described in \citet{Homolya2017} when most
  optimisations were done in COFFEE~\cite{Luporini2015,Luporini2016}.
\item \texttt{tensor} mode has two unique passes:
  \begin{enumerate}[label={(\arabic*)}]\addtocounter{enumi}{6}
  \item Flatten \texttt{Concatenate} nodes, destroying their inner
    structure.
  \item Attempt to refactorise the integrand expression such that
    quadrature is pre-evaluated at compile time.  This mimics the 
    older \emph{tensor representation} in
    FFC~\cite{Kirby2006,Kirby2007}, which is highly-performant for low-degree,
    constant-coefficient bilinear forms on affine simplices.
  \end{enumerate}
\end{itemize}
We must also note that modes have no effect on the evaluation of
parametrising functions.  For this purpose, we always apply passes (1)
and (3), as well as sum factorisation according to
\cref{sec:sum_fact_coeff}.

\section{Evaluation}
\label{sec:evaluation}

We now experimentally evaluate the performance benefits of this work,
using the new \texttt{spectral} mode of TSFC with all FInAT elements.
For comparison, we take a FIAT/\texttt{coffee} mode as baseline:
\begin{itemize}
\item Most FInAT elements are disabled, and instead we use the
  corresponding FIAT elements through the generic wrapper
  (\cref{sec:fiat_wrapper}).  Vector and tensor elements
  (\cref{sec:tensor_element}) are an exception: since we simplified
  the kernel interface with the FInAT transition, these elements have
  no direct equivalent in FIAT.
\item The transformations of the \texttt{coffee} mode are applied.
  Kronecker delta nodes are replaced with indexing of an identity
  matrix.
\end{itemize}

We consider a range of polynomial degrees, and in most cases we plot
the ratio of the number of degrees of freedom (DoFs) and execution time.
This metric helps to compare the relative cost of different polynomial
degrees.

The experiments were run on an workstation with two 2.6 GHz, 8-core
E5-2640~v3 (Haswell) CPUs, for a total of 16 cores.  The generated C
kernels were compiled with \texttt{-march=native -O3 -ffast-math}
using GCC~5.4.0 provided by Ubuntu 16.04.3 LTS.

\subsection{Matrix assembly}

\begin{figure}
  \centering
  \begin{tikzpicture}
    \begin{loglogaxis}[
      legend cell align=left,
      legend pos=outer north east,
      xlabel={Polynomial degree ($ \mathrm{n} $)},
      xtick={1,2,3,4,5,6,7},
      xticklabels={1,2,3,4,5,6,7},
      ylabel={DoFs/$s$},
      ]
      \addplot[dotted,mark=o,mark options={solid}] table
      [x=degree, y=rate, col sep=comma] {data/stokes_momentum_base.csv};
      \addlegendentry{\texttt{coffee}}

      \addplot[dotted,mark=*,mark options={solid,fill}] table
      [x=degree, y=rate, col sep=comma] {data/stokes_momentum_spectral.csv};
      \addlegendentry{\texttt{spectral}}
    \end{loglogaxis}
  \end{tikzpicture}
  \caption{Stokes momentum term assembly on a tetrahedral mesh,
    excluding overheads like sparsity pattern creation and matrix
    initialisation.}
  \label{fig:stokes_momentum}
\end{figure}
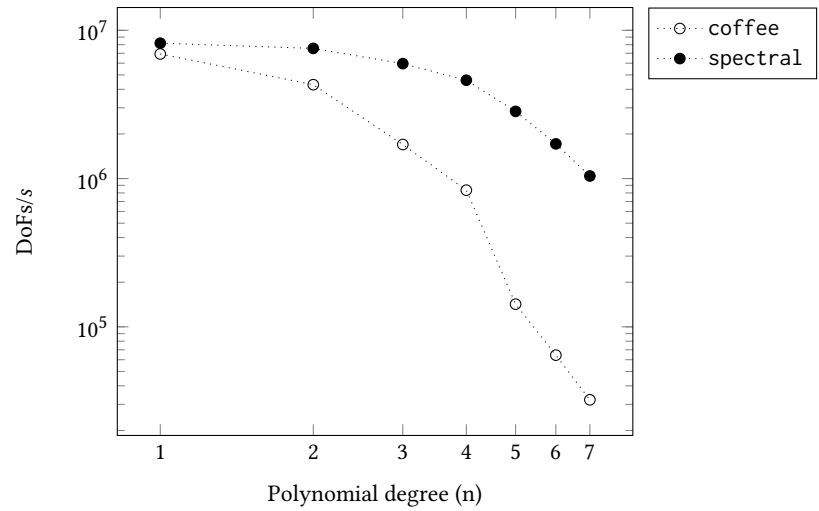

To demonstrate delta cancellation, we first consider the Stokes
momentum term
\begin{equation}
  \int_\Omega \nabla u : \nabla v \dx
\end{equation}
where $ u $ and $ v $ are vector-valued trial and test functions
respectively.  To rule out the additional effects of sum
factorisation, we evaluate this form on a tetrahedral mesh.
\Cref{fig:stokes_momentum} shows the difference in performance across
a range of polynomial degrees, with more speed up for higher degrees.
Note that for matrix assembly, if assembling an $ N \times N $ sparse
matrix takes $ t $ seconds, then the DoFs/$s$ rate is $ N / t $.

Since some degrees of freedom are shared between cells, assembly needs
to work on those DoFs multiple times.  This effect reduces the
DoFs/$s$ rate, and is stronger for low polynomial degrees.  To
approximately quantify this, we consider $ Q_n $ elements:
the number of unique degrees of freedom -- neglecting boundary effects
-- is exactly $ n^d $ per cell, while the number of degrees of freedom
is $ (n+1)^d $ for each cell.  Suppose $ O(n^r) $ is the required number
of floating-point operations per cell, where $ r $ is the appropriate
power.  For low polynomial degrees, $ c \cdot (n+1)^r $ with some
constant $ c $ is generally a closer approximation than $ c \cdot n^r $.  Therefore,
assuming a degree-independent FLOPS rate, the DoFs/$s$ measure is
approximated as
\begin{equation}
  C \cdot n^d / (n+1)^r .
\end{equation}

Next, we consider the Laplace operator
$ \int_\Omega \nabla u \cdot \nabla v \dx $ on a hexahedral mesh to
demonstrate sum factorisation.  The expected per-cell algorithmic
complexity is $ O(n^9) $ without and $ O(n^7) $ with sum
factorisation, as anticipated in \cref{tab:form_ordo}.
\Cref{fig:poisson_assembly} shows measurement data and confirms these
expectations.  We see orders of magnitude difference in performance
for the highest degrees.

\newcommand{\fitCurve}[4]{%
  \pgfplotstablegetelem{#2}{C}\of#1
  \pgfmathsetmacro{#3}{\pgfplotsretval}
  \pgfplotstablegetelem{#2}{a}\of#1
  \pgfmathsetmacro{\a}{\pgfplotsretval}
  \pgfplotstablegetelem{#2}{b}\of#1
  \pgfmathsetmacro{\b}{\pgfplotsretval}
  \addplot[densely dashdotted,domain=\a:\b] {#4};
}

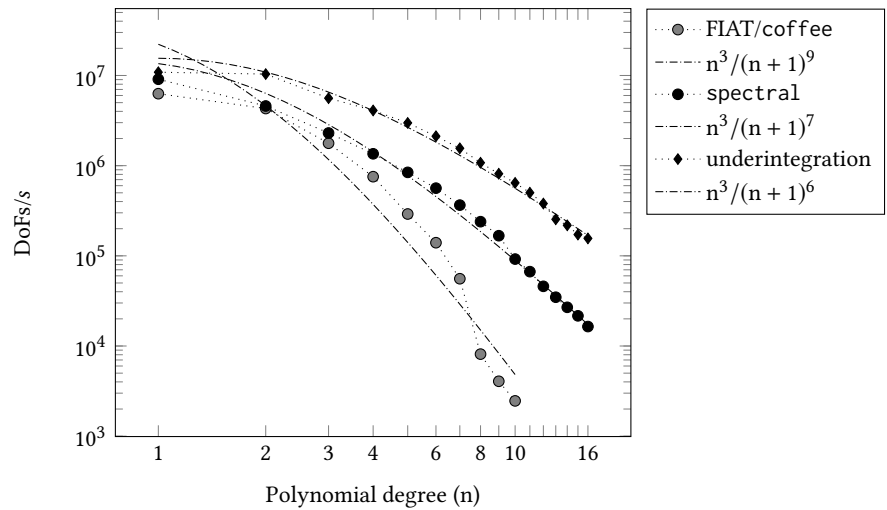
\begin{figure}
  \centering
  \begin{tikzpicture}
    \begin{loglogaxis}[
      legend cell align=left,
      legend pos=outer north east,
      xlabel={Polynomial degree ($ \mathrm{n} $)},
      xtick={1,2,3,4,5,6,7,8,9,10,11,12,13,14,15,16},
      xticklabels={1,2,3,4,,6,,8,,10,,,,,,16},
      ylabel={DoFs/$s$},
      ]
      \addplot[dotted,mark=*,mark options={solid,fill=gray}] table
      [x=degree, y=rate, col sep=comma] {data/bipoisson_base.csv};
      \addlegendentry{FIAT/\texttt{coffee}}

      \pgfplotstableread{data/bipoisson.dat}{\data}
      \fitCurve{\data}{0}{\C}{\C * x^3 / (x+1)^9}
      \addlegendentry{$ \mathrm{n^3 / (n+1)^9} $}

      \addplot[dotted,mark=*,mark options={solid,fill}] table
      [x=degree, y=rate, col sep=comma] {data/bipoisson_spectral.csv};
      \addlegendentry{\texttt{spectral}}

      \fitCurve{\data}{1}{\C}{\C * x^3 / (x+1)^7}
      \addlegendentry{$ \mathrm{n^3 / (n+1)^7} $}

      \addplot[dotted,mark=diamond*,mark options={solid,fill}] table
      [x=degree, y=rate, col sep=comma] {data/bipoisson_underintegration.csv};
      \addlegendentry{underintegration}

      \fitCurve{\data}{2}{\C}{\C * x^3 / (x+1)^6}
      \addlegendentry{$ \mathrm{n^3 / (n+1)^6} $}
    \end{loglogaxis}
  \end{tikzpicture}
  \caption{Number of degrees of freedom calculated per second for
    hexahedral Laplace operator \emph{assembly}, excluding overheads
    like sparsity pattern creation and matrix initialisation.}
  \label{fig:poisson_assembly}
\end{figure}

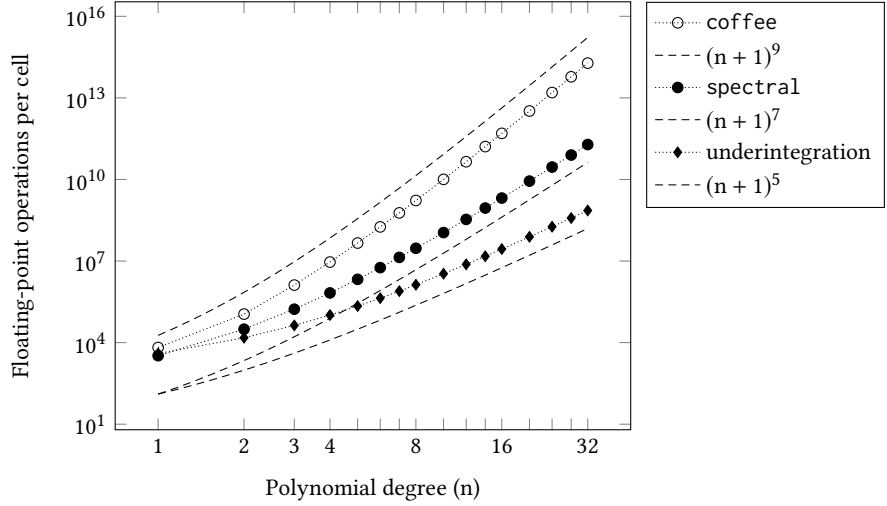
\begin{figure}
  \centering
  \begin{tikzpicture}
    \begin{loglogaxis}[
      legend cell align=left,
      legend pos=outer north east,
      xlabel={Polynomial degree ($ \mathrm{n} $)},
      xtick={1,2,3,4,5,6,7,8,10,12,14,16,20,24,28,32},
      xticklabels={1,2,3,4,,,,8,,,,16,,,,32},
      ylabel={Floating-point operations per cell},
      ]
      \addplot[densely dotted,mark=o,mark options={solid}]
      table [x=degree, y=coffee, col sep=comma] {data/bipoisson_flops.csv};
      \addlegendentry{\texttt{coffee}}

      \addplot[densely dashed,domain=1:32] {36*(x+1)^9};
      \addlegendentry{$ \mathrm{(n+1)^9} $}

      \addplot[densely dotted,mark=*,mark options={solid}]
      table [x=degree, y=spectral, col sep=comma] {data/bipoisson_flops.csv};
      \addlegendentry{\texttt{spectral}}

      \addplot[densely dashed,domain=1:32] {(x+1)^7};
      \addlegendentry{$ \mathrm{(n+1)^7} $}

      \addplot[densely dotted,mark=diamond*,mark options={solid}]
      table [x=degree, y=gll_spectral, col sep=comma] {data/bipoisson_flops.csv};
      \addlegendentry{underintegration}

      \addplot[densely dashed,domain=1:32] {4*(x+1)^5};
      \addlegendentry{$ \mathrm{(n+1)^5} $}
    \end{loglogaxis}
  \end{tikzpicture}
  \caption{Floating-point operations in generated kernels for
    hexahedral Laplace operator assembly.}
  \label{fig:poisson_flops}
\end{figure}

We now demonstrate an interesting combination of delta cancellation
and sum factorisation.  Still considering the Laplace operator, we take
the test and trial functions from a Gauss--Lobatto--Legendre finite
element basis, and we consider a quadrature rule with collocated
quadrature points (\cref{sec:colocated}).  This quadrature rule is
insufficient for exact integration, so the user has to specify the
quadrature rule manually to enable this optimisation.  This case
appears in \cref{fig:poisson_assembly} as ``underintegration''.
However, the collocation enables FInAT to construct the substitution
\begin{equation}
  \frac{\partial \Psi_i}{\partial \hat{x}_k}(\xi_q) \mapsto
  \begin{bmatrix}
    D\mathbf{\Psi}^{(1)}_{i_1 q_1} \delta_{i_2 q_2} \delta_{i_3 q_3} \\
    \delta_{i_1 q_1} D\mathbf{\Psi}^{(2)}_{i_2 q_2} \delta_{i_3 q_3} \\
    \delta_{i_1 q_1} \delta_{i_2 q_2} D\mathbf{\Psi}^{(3)}_{i_3 q_3}
  \end{bmatrix}_k
\end{equation}
where $ i = (i_1, i_2, i_3) $ and $ q = (q_1, q_2, q_3) $, and
$ D\mathbf{\Psi}^{(1)} $, $ D\mathbf{\Psi}^{(2)} $, and
$ D\mathbf{\Psi}^{(3)} $ are tabulation matrices for the derivative of
the interval elements.  The algorithmic complexity of assembling the
Laplace operator on nonaffine $ d $-cubes could thus be reduced to
$ O(n^{d + 2}) $, while sum factorisation alone gets $ O(n^{2 d + 1}) $.
Therefore, we would expect ``underintegration'' to approach $ O(n^5) $
on a hexahedral mesh, but in \cref{fig:poisson_assembly} it approaches
$ O(n^6) $ instead, since \emph{global assembly} in Firedrake still
assumes a \emph{dense} element stiffness matrix.  However,
analytically calculating the number of floating-point operations in
the generated kernels, we can confirm that FInAT and TSFC optimise
this case correctly (see \cref{fig:poisson_flops}).

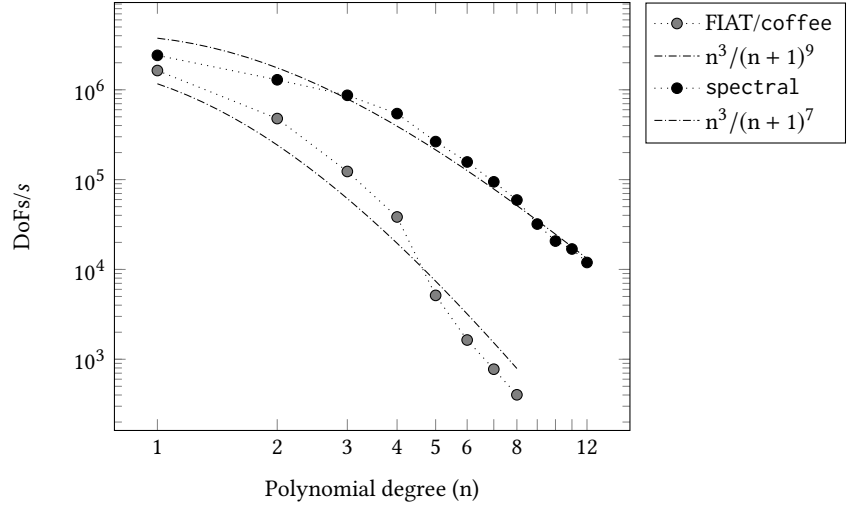
\begin{figure}
  \centering
  \begin{tikzpicture}
    \begin{loglogaxis}[
      legend cell align=left,
      legend pos=outer north east,
      xlabel={Polynomial degree ($ \mathrm{n} $)},
      xtick={1,2,3,4,5,6,7,8,9,10,11,12},
      xticklabels={1,2,3,4,5,6,,8,,,,12},
      ylabel={DoFs/$s$},
      ]
      \addplot[dotted,mark=*,mark options={solid,fill=gray}] table
      [x=degree, y=rate, col sep=comma] {data/bihyperelastic_base.csv};
      \addlegendentry{FIAT/\texttt{coffee}}

      \pgfplotstableread{data/bihyperelastic.dat}{\data}
      \fitCurve{\data}{0}{\C}{\C * x^3 / (x+1)^9}
      \addlegendentry{$ \mathrm{n^3 / (n+1)^9} $}

      \addplot[dotted,mark=*,mark options={solid,fill}] table
      [x=degree, y=rate, col sep=comma] {data/bihyperelastic_spectral.csv};
      \addlegendentry{\texttt{spectral}}

      \fitCurve{\data}{1}{\C}{\C * x^3 / (x+1)^7}
      \addlegendentry{$ \mathrm{n^3 / (n+1)^7} $}
    \end{loglogaxis}
  \end{tikzpicture}
  \caption{Number of degrees of freedom calculated per second for the
    left-hand side \emph{assembly} of a hexahedral hyperelastic model,
    excluding overheads like sparsity pattern creation and matrix
    initialisation.}
  \label{fig:hyperelasticity_assembly}
\end{figure}

Of course, the objective of this work is not simply to have another
attempt at optimising the Laplace operator, but to have an automatic
code generation system that can carry out these optimisations in
principle on any form.  As a more complicated example, we consider the
simplest hyperelastic material model, the Saint~Venant--Kirchhoff
model \cite[p.~529]{logg2012automated}.  First, we define the strain
energy function over the displacement vector field $ \mathbf{u} $:
\begin{align}
  \mathbf{F} &= \mathbf{I} + \nabla \mathbf{u}
  && \triangleright \text{Deformation gradient} \\
  \mathbf{C} &= \mathbf{F}^T \mathbf{F}
  && \triangleright \text{Right Cauchy-Green tensor}  \\
  \mathbf{E} &= (\mathbf{C} - \mathbf{I}) / 2
  && \triangleright \text{Euler-Lagrange strain tensor}  \\
  \Psi &= \frac{\lambda}{2}[\tr(\mathbf{E})]^2 + \mu \tr(\mathbf{E}^2)
  && \triangleright \text{Strain energy function}
\end{align}
where $ \lambda $ and $ \mu $ are the \emph{Lamé parameters}, and
$ \mathbf{I} $ is the identity matrix.  Now, we define the
Piola-Kirchhoff stress tensors:
\begin{align}
  \mathbf{S} &= \frac{\partial \Psi}{\partial \mathbf{E}}
  && \triangleright \text{Second Piola-Kirchhoff stress tensor} \\
  \mathbf{P} &= \mathbf{F} \mathbf{S}
  && \triangleright \text{First Piola-Kirchhoff stress tensor}
\end{align}
UFL derives automatically that
$ \mathbf{S} = \lambda \tr(\mathbf{E}) \mathbf{I} + 2 \mu \mathbf{E}
$.  Finally, the residual form of this nonlinear problem is
\begin{equation}
  r = \mathbf{P} : \nabla \mathbf{v} - \mathbf{b} \cdot \mathbf{v}
\end{equation}
where $ \mathbf{b} $ is the external forcing.  To assemble a left-hand
side, one must linearise the residual around an \emph{approximate}
solution $ \mathbf{u} $:
\begin{equation}
  \label{eq:bilinear_form}
  a = \delta r(\mathbf{u}; \delta \mathbf{u}) = \lim_{\epsilon \to 0}
  \frac{r(\mathbf{u} + \epsilon \delta \mathbf{u}) -
    r(\mathbf{u})}{\epsilon}
\end{equation}
This bilinear form has trial function $ \delta \mathbf{u} $, test
function $ \mathbf{v} $, and $ \mathbf{u} $ is a coefficient of the
form.  \Cref{fig:hyperelasticity_assembly} plots the performance as a
function of polynomial degree.

\begin{figure}
  \centering
  \begin{tikzpicture}
    \begin{loglogaxis}[
      legend cell align=left,
      legend pos=outer north east,
      xlabel={Polynomial degree ($ \mathrm{n} $)},
      xtick={1,2,3,4,5,6,7,8,9,10,11,12},
      xticklabels={1,2,3,4,5,6,,8,,,,12},
      ylabel={DoFs/$s$},
      ]
      \addplot[dotted,mark=*,mark options={solid,fill=gray}] table
      [x=degree, y=rate, col sep=comma] {data/bicurlcurl_base.csv};
      \addlegendentry{FIAT/\texttt{coffee}}

      \pgfplotstableread{data/bicurlcurl.dat}{\data}
      \fitCurve{\data}{0}{\C}{\C * x^3 / (x+1)^9}
      \addlegendentry{$ \mathrm{n^3 / (n+1)^9} $}

      \addplot[dotted,mark=*,mark options={solid,fill}] table
      [x=degree, y=rate, col sep=comma] {data/bicurlcurl_spectral.csv};
      \addlegendentry{\texttt{spectral}}

      \fitCurve{\data}{1}{\C}{\C * x^3 / (x+1)^7}
      \addlegendentry{$ \mathrm{n^3 / (n+1)^7} $}
    \end{loglogaxis}
  \end{tikzpicture}
  \caption{Number of degrees of freedom calculated per second for
    hexahedral curl-curl operator \emph{assembly}, excluding overheads
    like sparsity pattern creation and matrix initialisation.}
  \label{fig:curl_curl_assembly}
\end{figure}
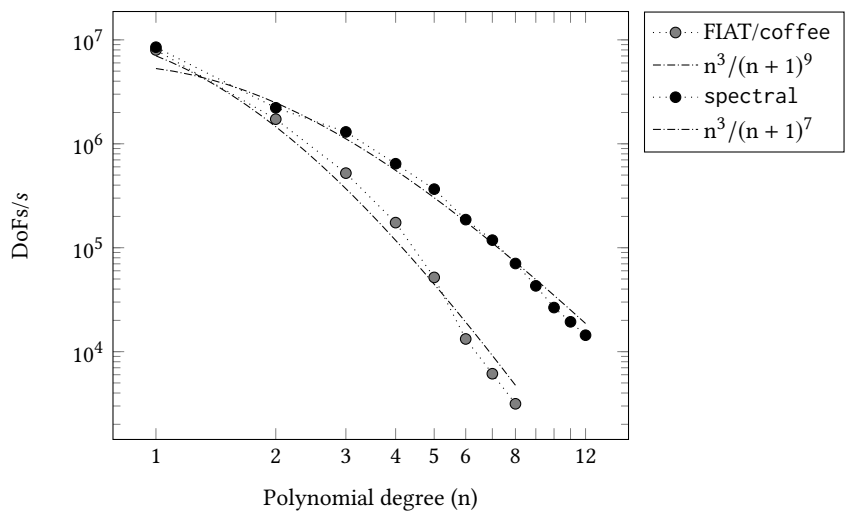

Finally, to demonstrate sum factorisation on \Hdiv{} and \Hcurl{}
conforming elements, we also consider the curl-curl operator, defined
as
\begin{equation}
  \label{eq:curl_curl_form}
  a(u, v) = \int_\Omega (\nabla \times u) \cdot (\nabla \times v) \dx .
\end{equation}
Here we use NCE elements, the hexahedral \Hcurl{} conforming element
in the $ \mathcal{Q}^- $ family.  \Cref{fig:curl_curl_assembly} shows
that sum factorisation also works for this finite element with the
\texttt{spectral} mode.

\subsection{Operator action and linear forms}

Since there are $ O(n^d) $ nonzero entries per row in the assembled
matrix, $ n $ being the polynomial order and $ d $ the dimension, the
matrix-free application of operator action is especially attractive
for higher-order discretisations.  The reasons are twofold.  First,
the memory requirement of storing the assembled matrix may be
prohibitively expensive.  In fact, this is the reason we only tested
matrix assembly for degees up to 16. At that point, the size of the
dense local tensor is 193MB, and we often need several temporaries of
similar size, all of which are allocated on the stack.  Even with
unlimited stack size and just one cell plus halo region per core, we
quickly run out of the available 64GB memory.  Second, fast finite element assembly with
sum factorisation will outperform matrix-vector multiplication with a
pre-assembled sparse matrix, even if the cost of matrix assembly is
ignored.  This is because sparse matrix-vector multiplication takes
$ O(n^{2 d}) $ time per cell, while matrix-free assembly of operator
action with sum factorisation only needs $ O(n^{d + 1}) $.

\begin{figure}
  \centering
  \begin{tikzpicture}
    \begin{loglogaxis}[
      legend cell align=left,
      legend pos=outer north east,
      xlabel={Polynomial degree ($ \mathrm{n} $)},
      xtick={1,2,3,4,5,6,7,8,10,12,14,16,20,24,28,32},
      xticklabels={1,2,3,4,,6,,8,,12,,16,,,,32},
      ylabel={DoFs/$s$},
      ]
      \addplot[densely dotted,mark=star,mark options={solid}] table
      [x=degree, y=rate, col sep=comma] {data/poisson_spmv.csv};
      \addlegendentry{\textsc{MatVec}}

      \addplot[dotted,mark=*,mark options={solid,fill=gray}] table
      [x=degree, y=rate, col sep=comma] {data/poisson_base.csv};
      \addlegendentry{FIAT/\texttt{coffee}}
      \pgfplotstableread{data/poisson.dat}{\data}
      \fitCurve{\data}{0}{\C}{\C * x^3 / (x+1)^6}
      \addlegendentry{$ \mathrm{n^3 / (n+1)^6} $}

      \addplot[dotted,mark=*,mark options={solid,fill}] table
      [x=degree, y=rate, col sep=comma] {data/poisson_spectral.csv};
      \addlegendentry{\texttt{spectral}}
      \fitCurve{\data}{1}{\C}{\C * x^3 / (x+1)^4}
      \addlegendentry{$ \mathrm{n^3 / (n+1)^4} $}
    \end{loglogaxis}
  \end{tikzpicture}
  \caption{Number of degrees of freedom calculated per second for
    hexahedral Laplace operator \emph{action}.}
  \label{fig:poisson_action}
\end{figure}
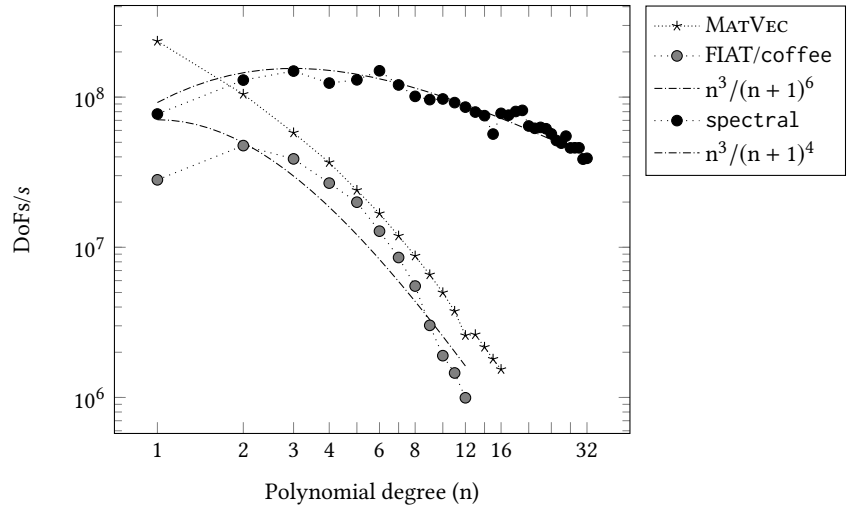

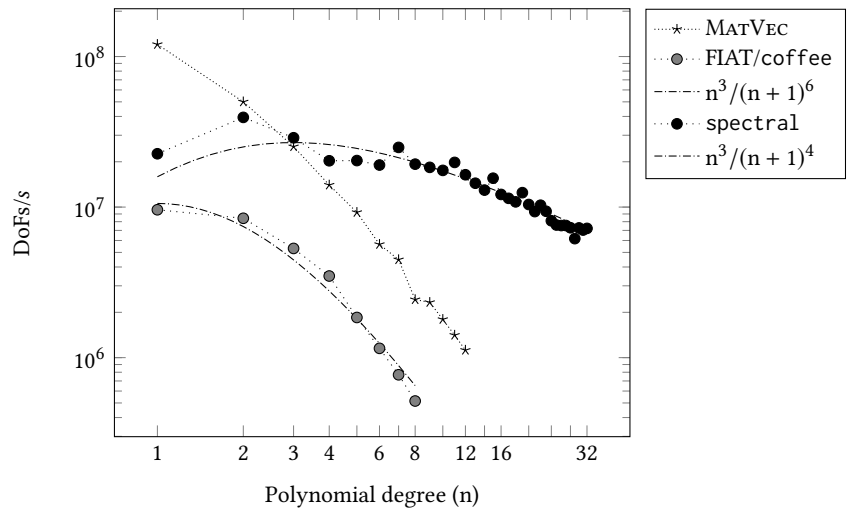
\begin{figure}
  \centering
  \begin{tikzpicture}
    \begin{loglogaxis}[
      legend cell align=left,
      legend pos=outer north east,
      xlabel={Polynomial degree ($ \mathrm{n} $)},
      xtick={1,2,3,4,5,6,7,8,10,12,14,16,20,24,28,32},
      xticklabels={1,2,3,4,,6,,8,,12,,16,,,,32},
      ylabel={DoFs/$s$},
      ]
      \addplot[densely dotted,mark=star,mark options={solid}] table
      [x=degree, y=rate, col sep=comma] {data/hyperelastic_spmv.csv};
      \addlegendentry{\textsc{MatVec}}

      \addplot[dotted,mark=*,mark options={solid,fill=gray}] table
      [x=degree, y=rate, col sep=comma] {data/hyperelastic_base.csv};
      \addlegendentry{FIAT/\texttt{coffee}}
      \pgfplotstableread{data/hyperelastic.dat}{\data}
      \fitCurve{\data}{0}{\C}{\C * x^3 / (x+1)^6}
      \addlegendentry{$ \mathrm{n^3 / (n+1)^6} $}

      \addplot[dotted,mark=*,mark options={solid,fill}] table
      [x=degree, y=rate, col sep=comma] {data/hyperelastic_spectral.csv};
      \addlegendentry{\texttt{spectral}}
      \fitCurve{\data}{1}{\C}{\C * x^3 / (x+1)^4}
      \addlegendentry{$ \mathrm{n^3 / (n+1)^4} $}
    \end{loglogaxis}
  \end{tikzpicture}
  \caption{Number of degrees of freedom calculated per second for the
    left-hand side \emph{action} of a hexahedral hyperelastic model.}
  \label{fig:hyperelasticity_action}
\end{figure}

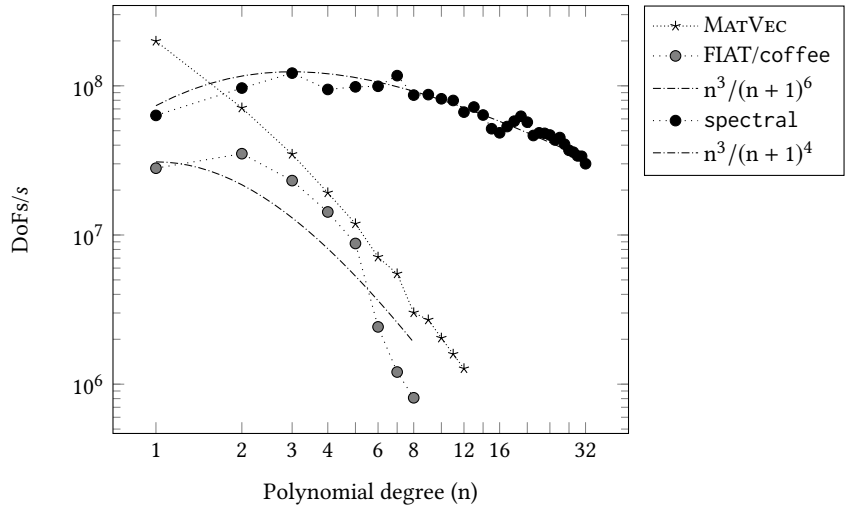
\begin{figure}
  \centering
  \begin{tikzpicture}
    \begin{loglogaxis}[
      legend cell align=left,
      legend pos=outer north east,
      xlabel={Polynomial degree ($ \mathrm{n} $)},
      xtick={1,2,3,4,5,6,7,8,10,12,14,16,20,24,28,32},
      xticklabels={1,2,3,4,,6,,8,,12,,16,,,,32},
      ylabel={DoFs/$s$},
      ]
      \addplot[densely dotted,mark=star,mark options={solid}] table
      [x=degree, y=rate, col sep=comma] {data/curlcurl_spmv.csv};
      \addlegendentry{\textsc{MatVec}}

      \addplot[dotted,mark=*,mark options={solid,fill=gray}] table
      [x=degree, y=rate, col sep=comma] {data/curlcurl_base.csv};
      \addlegendentry{FIAT/\texttt{coffee}}
      \pgfplotstableread{data/curlcurl.dat}{\data}
      \fitCurve{\data}{0}{\C}{\C * x^3 / (x+1)^6}
      \addlegendentry{$ \mathrm{n^3 / (n+1)^6} $}

      \addplot[dotted,mark=*,mark options={solid,fill}] table
      [x=degree, y=rate, col sep=comma] {data/curlcurl_spectral.csv};
      \addlegendentry{\texttt{spectral}}
      \fitCurve{\data}{1}{\C}{\C * x^3 / (x+1)^4}
      \addlegendentry{$ \mathrm{n^3 / (n+1)^4} $}
    \end{loglogaxis}
  \end{tikzpicture}
  \caption{Number of degrees of freedom calculated per second for
    hexahedral curl-curl operator \emph{action}.}
  \label{fig:curl_curl_action}
\end{figure}

We can directly compare sparse matrix-vector multiplication with
various approaches to matrix-free algorithms using the metric of
degrees of freedom per second, or DoFs/$s$.  This is plotted for
three configurations for calculating the action of the Laplace
operator in \cref{fig:poisson_action}, the left-hand side of a
hyperelastic model in \cref{fig:hyperelasticity_action}, and the
curl-curl operator in \cref{fig:curl_curl_action}.

\citet[\S 5.1]{Kirby2017} perform a similar comparison of matrix-free
actions to assembled PETSc~\cite{petsc-user-ref} matrices, but without
access to sum factorisation.  They find that:
\begin{enumerate}
\item Matrix-free applications are generally an $ O(1) $ factor slower
  than matrix-vector products.  When the Krylov subspace method runs
  no more than a few iterations, and the assembled matrix is not
  otherwise needed, then matrix-free applications could be overall
  cheaper the need for costly matrix assembly is eliminated.
  \label{itm:no_assembly}
\item For high polynomial orders, the memory requirement of the
  assembled matrix may simply be prohibitive.
\end{enumerate}
Using sum factorised assembly, however, matrix-free applications are
the clear fastest choice for high enough polynomial orders.  For low
orders, one must fall back to the considerations of point
\ref{itm:no_assembly}.

Note that $ n^3 / (n + 1)^4 $ initially increases before approaching
$ O(n^{-1}) $ as $ n \to \infty $, so the cost of sum factorised
matrix-free application initially decreases with increasing polynomial
order.  The same applies for $ n^2 / (n + 1)^3 $ in 2D.  As previously
discussed, this is due to the decreasing portion of degrees of freedom
that belong to multiple cells.  At the same time, this effect renders
sparse matrix-vector multiplication especially efficient at low
polynomial orders, since the contributions of different cells are
aggregated during matrix assembly.

We also observe that \texttt{coffee} mode does not outperform
\texttt{spectral} mode even in case of low polynomial orders.  This
makes the latter a good default choice in TSFC for all cases.
Furthermore, the use of FIAT elements was found to be infeasible in
the high-order regime, because it resulted in kernels that are over
100MB both as C code and as executable binaries.  Most of that size is
occupied by tabulation matrices, which grow $ O(n^{2 d}) $ in size,
nevertheless GCC may not finish compiling them within an hour.

Finally, a note on form compilation time.  Argument factorisation has
a noticeable, but not prohibitive cost, while all other described
transformations are cheap.  A complicated hyperelastic model, the
Holzapfel--Ogden model~\cite{Balaban2016} has been used for stress
testing form compilers \cite[\S 6]{Homolya2017}, since practical
finite element models are seldom more difficult to compile.  TSFC
compiles its left-hand side in 1.4 seconds in \texttt{vanilla} mode,
5.6 seconds in \texttt{coffee} mode, and 5.9 seconds in
\texttt{spectral} mode.

\section{Conclusion and future work}
\label{sec:conclusion}

We have presented a new, smarter library of finite elements, FInAT, which
is able to express the structure inherent to some finite elements.  We
described the implemented FInAT elements, as well as the form compiler
algorithms -- which were implemented in TSFC -- that exploit the
exposed structure.  With FInAT and TSFC, one can just write the weak
form in UFL, and automatically get:
\begin{itemize}
\item sum factorisation with continuous, discontinuous, \Hdiv{} and
  \Hcurl{} conforming elements on cuboid cells;
\item optimised evaluation at collocated quadrature points with
  underintegration is requested; and
\item minor optimisations with vector and tensor elements.
\end{itemize}
These techniques were known and have been applied in hand-written
numerical software before, however, we are now able to utilise these
optimisations in an automatic code generation setting in Firedrake.

This work is especially useful in combination with matrix-free methods
\cite{Kirby2017}, enabling the development of efficient high-order
numerical schemes while retaining high productivity.  Our measurements
show that on modern hardware one can increase the polynomial degree up
to 8--10 without a notable increase in the run time of matrix-free
operator applications (per degree of freedom).

Future work may include sum factorisation on simplicial cells, for
example, through Bernstein polynomials
\cite{Ainsworth2011,kirby2012fast,kirby2014low}, as well as the evaluation and
potentially optimisation of low-level performance on modern hardware.

\appendix

\section{Code availability}

For the sake of reproducibility, we have archived the specific
versions of Firedrake components on Zenodo that were used for these
measurements: PETSc~\citeyear{barry_smith_2017_1022071},
petsc4py~\citeyear{lisandro_dalcin_2017_1022068},
COFFEE~\citeyear{fabio_luporini_2017_573267},
PyOP2~\citeyear{florian_rathgeber_2017_1039612},
FIAT~\citeyear{miklos_homolya_2017_1022075},
FInAT~\citeyear{david_a_ham_2017_1039605},
UFL~\citeyear{anders_logg_2017_1022069},
TSFC~\citeyear{miklos_homolya_2017_1022066}, and
Firedrake~\citeyear{lawrence_mitchell_2017_1039613}.  For the
FIAT/\texttt{coffee} mode, we applied a custom patch to
TSFC~\citeyear{miklos_homolya_2017_1039640}.  The experimentation
framework is available at \cite{miklos_homolya_2017_1041785}.

\bibliographystyle{ACM-Reference-Format}
\bibliography{references,zenodo}

\end{document}